\documentclass[12pt,letterpaper]{article}

\usepackage{amsmath,amssymb}
\usepackage{graphicx}
\usepackage[utf8]{inputenc}
\usepackage{placeins}
\usepackage{caption}
\usepackage{subcaption}
\usepackage{geometry}
\usepackage{upgreek}
\usepackage{multirow}

\usepackage[style=numeric-comp,sorting=none,maxnames=4]{biblatex}
\usepackage{amsmath,amssymb}
\usepackage{bm}

\DeclareMathOperator{\E}{\mathbb{E}}
\newcommand{\Nor}{\mathcal{N}}
\newcommand{\Corr}{{\rm Corr}}

\newcommand{\Var}{{\rm Var}}

\newcommand{\Vh}{V^{1/2}}
\newcommand{\Uh}{U^{1/2}}

\newcommand{\DTW}{\$TW{}}
\DeclareMathOperator{\argmin}{argmin}

\title{NYSE Price Correlations Are Abitrageable\\
Over Hours and Predictable Over Years}
\author{William H. Press\\
Oden Institute for Computational Engineering and Sciences\\
The University of Texas at Austin}

\begin{document}
\maketitle

\begin{abstract}
Trade prices of about 1000 New York Stock Exchange-listed stocks are studied at one-minute time resolution over the continuous five year period 2018--2022. For each stock, in dollar-volume-weighted transaction time, the discrepancy from a Brownian-motion martingale is measured on timescales of minutes to several days. The result is well fit by a power-law shot-noise (or Gaussian) process with Hurst exponent 0.465, that is, slightly mean-reverting. As a check, we execute an arbitrage strategy on simulated Hurst-exponent data, and a comparable strategy in backtesting on the actual data, obtaining similar results (annualized returns $\sim 60$\% if zero transaction costs). Next examining the cross-correlation structure of the $\sim 1000$ stocks, we find that, counterintuitively, correlations increase with time lag in the range studied. We show that this behavior that can be quantitatively explained if the mean-reverting Hurst component of each stock is uncorrelated, i.e., does not share that stock's overall correlation with other stocks. Overall, we find that $\approx 45$\% of a stock's 1-hour returns variance is explained by its particular correlations to other stocks, but that most of this is simply explained by the movement of all stocks together. Unexpectedly, the fraction of variance explained is greatest when price volatility is high, for example during COVID-19 year 2020. An arbitrage strategy with cross-correlations does significantly better than without (annualized returns $\sim 100$\% if zero transaction costs). Measured correlations from any single year in 2018--2022 are about equally good in predicting all the other years, indicating that an overall correlation structure is persistent over the whole period.
\end{abstract}

\section{Introduction}
\label{sec1}

Only recently has tick-by-tick historical trading data across
whole markets like the New York Stock Exchange (NYSE) become available to all comers, at low cost and outside of proprietary settings. The sites Finnhub.io \cite{finn} and Polygon.io \cite{poly} are among current examples of such sources in a rapidly evolving landscape. Also only recently has GPU software like PyTorch \cite{pytorch} or TensorFlow \cite{tensorflow} made easy the exploitation of financial data sets at gigabyte and terabyte scales with desktop resources. Given these trends, one expects to see a new wave of published large-scale statistical studies of the behavior of markets. This paper is one such.

We study the collective behavior of five years (2018--2022) of $\sim$1000 NYSE listed stocks with continuous trading data at one minute resolution. The span of years is chosen to include year 2020, during which the COVID-19 pandemic roiled markets. Our interest is one with a long history, albeit somewhat neglected in recent years: to characterize and quantify deviations of the market from its so-called stylized facts, especially random-walk models, without introducing too much extra theoretical machinery; and to quantify the predictive power (in several respects, to be defined below) of the market's correlational structure. That the market deviates from its stylized facts is in no sense controversial \cite{Mandelbrot1971,lo1988,debondt1985,poterba1988,boudoukh1994,Baillie}. Our goal is new quantification in a carefully controlled large data set.

In \S\ref{secprelim} we summarize a set of these so-called facts, also reviewing the history of discovery of the necessity of a transaction time different from clock time. We describe the data set used in this paper. A technical point, we discuss the relation between returns computed as two-point instantaneous price differences, versus the difference of time-averaged prices. In \S\ref{sec2} we compute variograms (equivalent in some but not all ways to autocorrelation functions) over the data, noting some apparently mean-regressing long-term memory on timescales minutes to days. In \S\ref{sechurst} we review the venerable model of shot-noise (or, in the limit, Gaussian) processes with power-law (Hurst exponent) variograms, also known as fractional Brownian motion. This provides a convenient platform for discussing the advantage of variance analysis over autocorrelation, and for demonstrating directly that profitable arbitrage is possible in principle for such models (except for the case of a perfect random walk).

In \S\ref{secarbit} we apply \S\ref{sechurst}'s arbitrage trading strategy to the actual stock-market data and find it to be profitable about as predicted by its observed Hurst exponent. This serves to remove any lingering doubts that the long-memory, mean-reverting behavior measured is genuine, not an artifact of (e.g.) a flawed mapping of trading time to clock time.

In \S\ref{sec6} we turn to the cross-correlational structure of the $\sim 1000$ NYSE stocks. Measuring the correlation with one-hour returns over five years produces an interesting ``atlas" of correlation diagrams (shown in Supplementary Information). Measuring correlation as a function of the time-resolution of returns produces a seeming anomaly, which we show to be quantitatively explainable by the same apparent long-term memory as seen in the variograms.

Section \ref{sec7} looks at ``leave-one-out" predictions, where we calculate what hourly (or other) return should be expected of a stock, given the hourly returns of all other stocks in the same hour. Then, in \S\ref{sec8}, we test an arbitrage strategy based on leave-one-out predictions, with apparently robust positive results.

Section \ref{sec9} is additional discussion.

\section{Preliminaries}
\label{secprelim}
\subsection{Stylized Facts}
In economics, stylized facts, so-called, are empirical observations that evidence broad principles without being necessarily exact in all cases \cite{wiki:sfact}. Since the work of Fama \cite{fama1970} and elaboration of the Efficient Market Hypothesis (EMH) from the 1960s \cite{Sam}, recitations of stylized facts about the time history of market prices and returns usually include these \cite{Hommes}:
\begin{itemize}
    \item Asset prices are nonstationary and do (some kind of) random walk.
    \item Sequential asset returns, i.e., price changes, are (close to) independent. More formally, price evolution is (close to a) Markov process, therefore memoryless.
    \item In liquid markets, arbitrage opportunities are (almost) nonexistent. Or, equivalently: The Markov process is a martingale.  
    \item Because the market responds to the sum of innumerable small news effects, the Central Limit Theorem should apply, and the time series are expected to be Gaussian. 
\end{itemize}

Taken together, these stylized facts imply a Brownian motion random walk (also termed a Wiener process) as the null-hypothesis model for financial time series---at any rate, the model to disprove with contradictory data \cite{Mandelbrot}.

Historically, the last of these stylized facts was an immediate embarrassment, because, the distribution of many types of asset returns, sampled at equally spaced times, is strongly non-Gaussian, with positive kurtosis and fat tails. Some exotic solutions were proposed, for example Mandelbrot's examination of so-called stable distributions that could be the sum of many small effects, yet not Gaussian \cite{Mandelbrot,Petropulu}. But, as first notably studied by Clark in 1973 \cite{Clark}, the path to rescuing near-memoryless near-Gaussianity lay instead in what has come to be another stylized fact: 
\begin{itemize}
    \item Trading volume and price volatility are positively correlated.
\end{itemize}

\subsection{Transaction Time and Variogram}
\label{secttvar}

Clark interpreted the apparent non-Gaussianity as evidence that the market's Gaussian process advanced not in clock time, but rather in a ``transaction time" that, he noted empirically, was something close to cumulative trading volume \cite{Clark}. (Lacking a standard terminology, transaction time is by now also known as economic time, business time, trading time, market time, and operational time \cite{RVF}.) In the transaction time of cumulative trading volume, the distribution of returns per fixed time is close to normal. That transaction time and clock time converge over periods of order a month, was already implicit in 1965 work by Fama \cite{fama1965}.

Clark's observation unleashed a veritable flood of work in search of an exact or universal transaction time, one in which the normality of returns could be raised to the status of a law of nature (see references in \cite{Murphy}). A great deal of effort was expended on so-called subordinated processes, a special class of mappings from clock time to transaction time \cite{Clark}. Later, Ane and Geman \cite{Ane} advocated consideration of a broader class of mappings, calling attention to the result of Monroe \cite{Monroe} that \emph{any} \hbox{(semi-)} martingale process could be rendered as exactly a memoryless Gaussian process by an appropriate choice of stochastic time. 

Important for this paper is the fact, true for any martingale \cite{RVF} and especially for a memoryless Gaussian process, that the variance of returns $r$ over a (transaction) time $\uptau$ must scale exactly as $\uptau$, because the variances of returns over successive independent intervals simply add,
\begin{equation}
    V(\uptau) \equiv \Var[r(\uptau)] \equiv \E_t[(p(t+\uptau)-p(t))^2] \propto \uptau
\label{variogram}
\end{equation}
where $\E$ denotes expectation value.
In contexts other than financial, the statistic $V(\uptau)$ in equation \eqref{variogram} is termed the {\em variogram} of the time series $p(t)$ \cite{Cressie}, and we adopt that terminology here. Variogram analysis here is essentially equivalent to the financial literature's ``variance ratio test" \cite{poterba1988,lo1988}, though the latter term is more used in the context of a significance test than a measurement over multiple values of $\uptau$.  

Suppose a time series of prices, expressed as a function of some transaction time, is found to violate equation \eqref{variogram}.  How can we distinguish between the hypothesis that it is not a martingale---hence allows an arbitrage opportunity---and the hypothesis that we simply have the wrong transaction time, poorly approximating Monroe's perfect one, or that there exists some other systematic flaw in the data? This will be the key issue in \S\ref{sec2}, below.

\subsection{NYSE One-Minute Data Set}
\label{sec1p2}
We downloaded from Finnhub.io all available one-minute candles \cite{candlestick} for NYSE-listed stocks in the years 2018--2022, comprising substantially all such listed stocks. For a given stock, a one-minute candle exists if that stock traded during that minute.
Each candle consists of a universal timestamp (in Unix seconds) marking the beginning of the minute, four prices (open, high, low, close), and a volume of shares traded. A typical year has about 252 trading days, amounting to 98,280 trading minutes. The more active listed stocks trade in most minutes. There are 1,091 stocks with available candles spanning all five years.

Candles do not reveal how many separate trades occur in their minute, nor the sign of the trade (buyer- vs.~seller-initiated). If there are equalities among the four prices, then fewer than four trades may be present---but more than four is also possible. For these reasons, we in all cases take the mean of open, high, low, close as the representative average price for that minute, referring to this as ``one-minute resolution". Prices are considered known only in minutes with a reported candle, otherwise unknown. That is, we never hold over a previous price or interpolate between known prices.

We define resolution time intervals $\uptau > 1$ minute (e.g., $\uptau = 1$ hour) by sorting candles according to their timestamp into consecutive, nonoverlapping bins of length $\uptau$. We assign to each bin the mean of its candles timestamps as a time, the mean of its candles prices as a price. If a bin contains no candles, its price is unknown.

Of particular interest will be the estimation of the variances of the returns $r(\uptau)$ implied by price time series as defined above, leading to estimates of the variogram $V(\uptau)$, the covariance matrix $\mathbf{C}$ of multiple such returns (multiple listed stocks), and their implied correlation matrix $\boldsymbol{\rho}$.
Returns are defined by the logarithmic price difference of two consecutive known prices with the same resolution $\uptau$. Associated with each return is its time interval, the difference of its two average times, generally not an integer multiple of $\uptau$. We calculate $\sigma^2$, $\mathbf{C}$, and $\boldsymbol{\rho}$ by the methods described in \cite{press2023}, which are designed for such cases of asynchronous sampling.

\subsection{Difference-of-Average vs.~Two-Point Difference}
\label{sec1p22}
The astute reader will have noticed that we have introduced two slightly different definitions for $r(\uptau)$, the return over time $\uptau$ whose variance is the variogram
$V(\uptau)$. Equation \eqref{variogram} defines $r(\uptau)$ as the difference of two
point (i.e., instantaneous) prices spaced apart by time $\uptau$. Section \ref{sec1p2}, on the other hand, implies a definition that first averages prices into bins of length $\uptau$ and then defines $r(\uptau)$ as the difference of adjacent bins. How are these related?

In Supplementary Information \ref{suppinfo11}, we show that, for the case of a memoryless Gaussian process, the difference-of-average method yields a result that differs by a constant factor 2/3 from the two-point difference method. Since no
results in this paper will depend on the the absolute normalization
of the returns, the factor is irrelevant. More important is to compare the fractional accuracy to which the variogram can be measured for two methods, with input data consisting of a fixed number of one-minute candles, as described above in \S\ref{sec1p2}. Supplementary Information \ref{suppinfo12} demonstrates that the accuracies are close to equal.
With the data as described, however, the two-point difference method shows a bias that increases as $\uptau$ decreases towards one minute, arising because the candle prices are not in fact point prices but themselves averages. The difference-of-average method shows less such bias and is henceforth the method of choice for this paper.

\begin{figure}[ht]
\centering
\includegraphics[width=10cm]{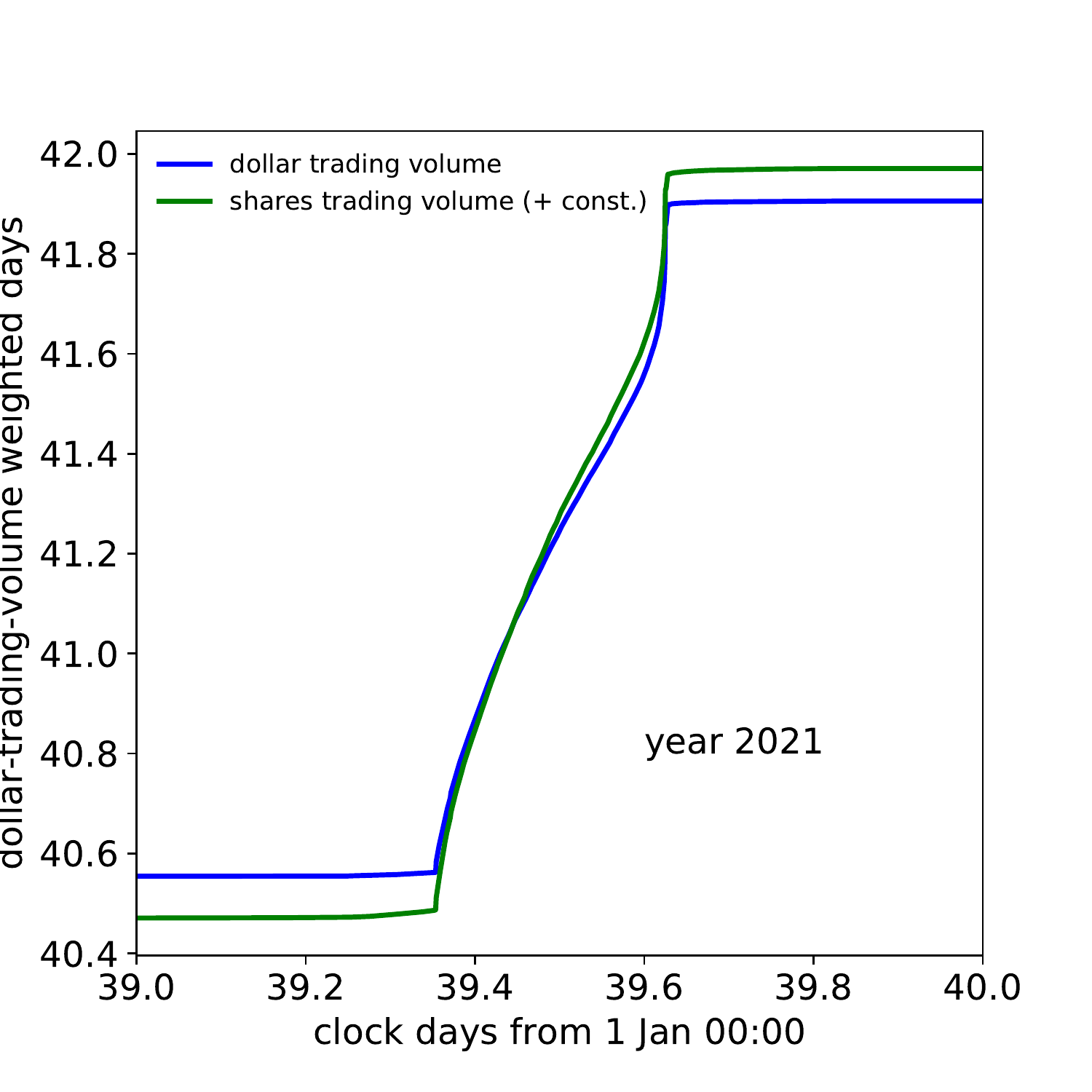}
\caption{Clock time versus dollar-trading-weighted and volume-trading-weighted times for a 24-hour clock interval in year 2021. The sharp rises at the beginning and end of the NYSE trading day are peaks in trading volume, including the opening and closing auctions. Almost no \DTW or VWT time elapses between trading sessions.}
\label{fig1}
\end{figure}

\begin{figure}[ht]
\centering
\includegraphics[width=16cm]{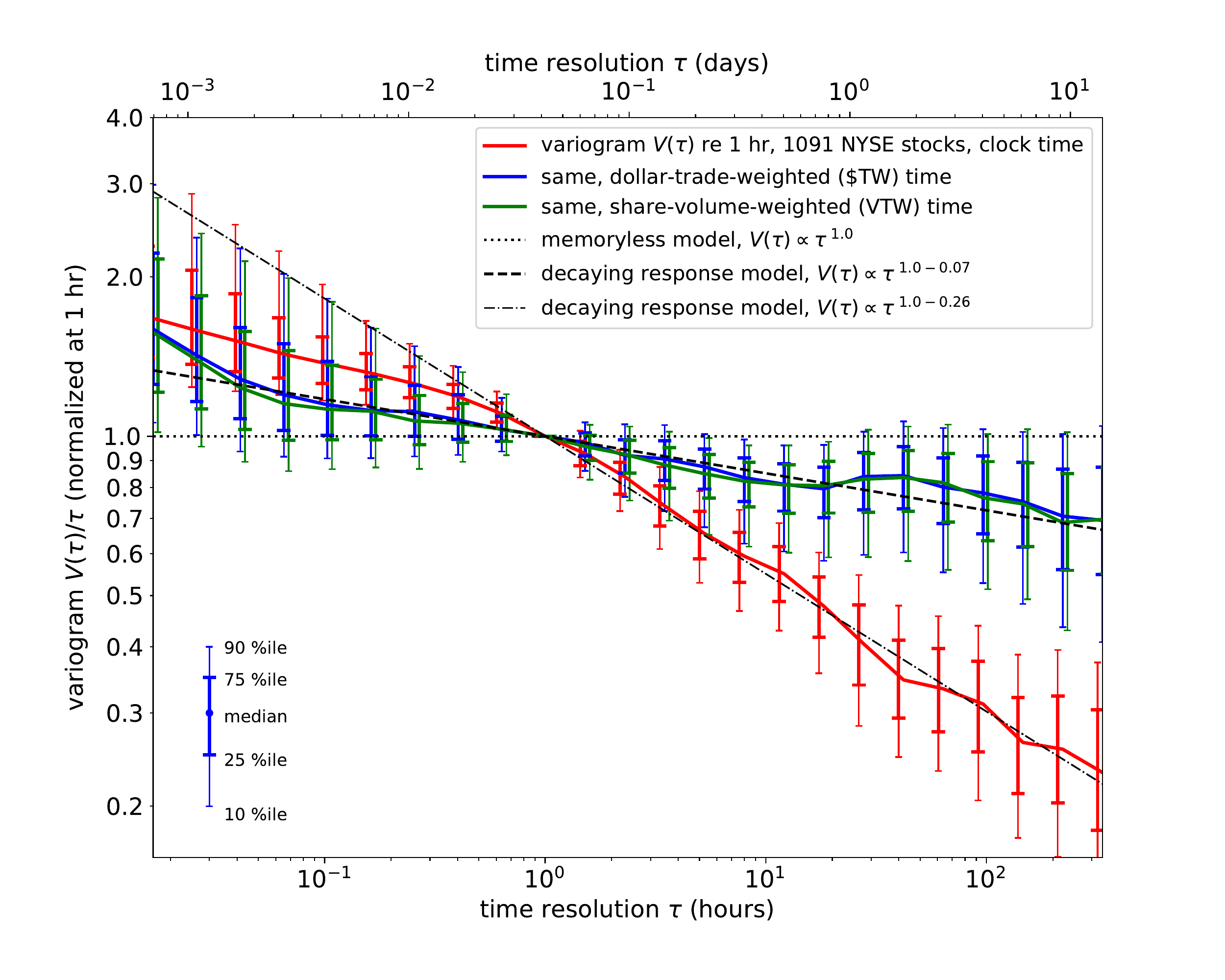}
\caption{Observed variograms of 1,091 NYSE stock prices as a function of resolution time interval $\uptau$ during year 2019. The ordinate is $V(\uptau)/\uptau$, so a memoryless random walk plots as constant. Both versions of transaction time, \$TW and VTW (see text), yield results well fit by a small negative exponent $-0.07$. Clock time (unlike the transaction times) shows a steeper slope and an artifactual break at $\sim 1$ hr.}
\label{fig2}
\end{figure}

\section{Variogram Estimate for NYSE Stocks}
\label{sec2}

Now making a choice among similar available alternatives, we adopt as a definition of transaction time cumulative New York Stock Exchange dollar-trading volume, normalized to clock time at one year. In other words, one dollar-trading-weighted hour (abbreviated \DTW hr) is an interval in which a fraction $1/8760$ of the year's dollar trading volume occurs ($1/8784$ in leap years). We can also compare \DTW results to their volume trading weighted counterparts, time unit abbreviated VTW hr, where the total share volume in a year is normalized to the number of hours in the year. As an example, Figure \ref{fig1} shows the time mapping of these definitions for calendar February 9, 2021, a typical day. In clock time, the shortest \DTW~minute is about 200 milliseconds (spanning a closing auction), the longest is about 3.6 days (spanning a holiday weekend). The shortest \DTW~hour is about 12 clock minutes (again, spanning a closing auction); the longest spans the same 3.6 day holiday.

\FloatBarrier

From the 2019 one-minute candle data on 1,091 stocks described in \S\ref{sec1p2}, we computed
separate variograms $V(\uptau)$ for each stock's prices, scaling each to have $V(1\text{ hr}) \equiv 1$. Figure \ref{fig2} shows the results as a function of resolution time interval $\uptau$. The ensemble of variograms is summarized in the figure as percentile sticks with markers at the 10, 25, 50, 75, and 90 percentiles. Displayed in this way, one sees that the dispersion of individual stocks around the median of all the stocks is surprisingly small, indicating an approximately ``universal" scaling law for $V(\uptau)$, both for individual stocks and for index averages.

The figure shows results for three different time parameterizations, clock time, \DTW, and VWT.
The ordinate $V(\uptau)/\uptau$ is chosen to make a memoryless random walk plot as a constant line.
First looking at the result for clock time (red curve and sticks), one sees a function decreasing monotonically, implying at first glance a process with a significant tendency towards reverting to the mean and, in any case, far from memoryless. However, we already know from the work cited in \S\ref{secttvar} above, that such effects can be produced by choosing a poor transaction-time parameterization. In particular, the change in logarithmic slope at $\sim 1$ hr is immediately suggestive of such an effect, as (e.g.) the intervals $\uptau$ come to straddle the beginning or end of trading days. So, we dismiss the seeming long-term memory of the red curve as an artifact.

Next turn attention to the blue and green curves and sticks, the same price data now mapped into either \DTW (blue) or VTW (green) transaction time. A first observation is that the blue and green results are virtually identical, implying that results are not sensitive to the exact choice of volume-weighted transaction time. A second observation is that the results are surprisingly well fit by a single logarithmic slope (power law), remarkably featureless from $\sim 4$ minutes to $\sim 1$ week, a range of more than three orders of magnitude. Summarizing, we see in the Figure approximately,
\begin{equation}
    V(\uptau) \propto \uptau^{\,1.0 - 0.07}, \qquad
    0.06\text{ hr} < \uptau < 200\text{ hr}
\label{eq2}
\end{equation}
that is, an exponent differing by $-0.07$ from a memoryless process.

We now face a dilemma: Should we believe (Hypothesis 1) that, as for the red curve,
the small exponent $-0.07$ is an artifact of having chosen a poor transaction time. In that case, a perfect Monroe stochastic clock \cite{Monroe} would yield a horizontal line and an implied memoryless random-walk process. Or (Hypothesis 2), should we believe that the small exponent is evidence of persistent-returns memory, here seen very generally across most stocks, with a timescale of minutes to days. In general, such a memory implies an (at least theoretical) arbitrage opportunity \cite{Mandelbrot1971,RVF}.

Hypothesis 2 is not ruled out a priori. There exist, after all, generally profitable quant hedge funds in equities markets \cite{Lo, Zuckerman, Quants}. Still, it does seem peculiar that a broad market inefficiency like this should survive in an exchange as liquid and visible as NYSE. We might ask it this result is an artifact of the year chosen, 2019. But (see Supplementary Information \S\ref{secfiveyear})
virtually identical results are seen for years 2018, 2019, 2021, and 2022. Only COVID-19 year 2020 shows some anomalies, but with still a small negative-exponent trend.

In the remainder of this paper, we take the view that the only really persuasive way to demonstrate the truth of Hypothesis 2 is to demonstrate, by backtesting on the actual price data, profitable arbitrage opportunity, and also to do this in the context of a controlled model that gives comparable results without the possibility of ``unknown unknowns".
The next section provides such a model.

\section{Hurst Power-Law Models}
\label{sechurst}

A shot-noise process \cite{Rice, Parzen} is a time series generated as the sum of impulses at Poisson-random times. Or, a slight generalization, it can be the sum of a normal random variable multiplying such impulses \cite{Lund},
\begin{equation}
    p(t) = \sum_i s_i f(t-t_i),\quad s_i \sim \text{Normal}(0,\sigma),
    \; (t_{i+1}-t_i) \sim \text{Exponential}(\lambda)
\label{eqshot}
\end{equation}
where $f(.)$ is the impulse response, $\sigma$ scales the impulses, and $\lambda$ is the
Poisson-process rate. (The difference between successive times in a Poisson process is,
of course, exponentially distributed.)

\begin{figure}[ht]
\centering
\includegraphics[width=12cm]{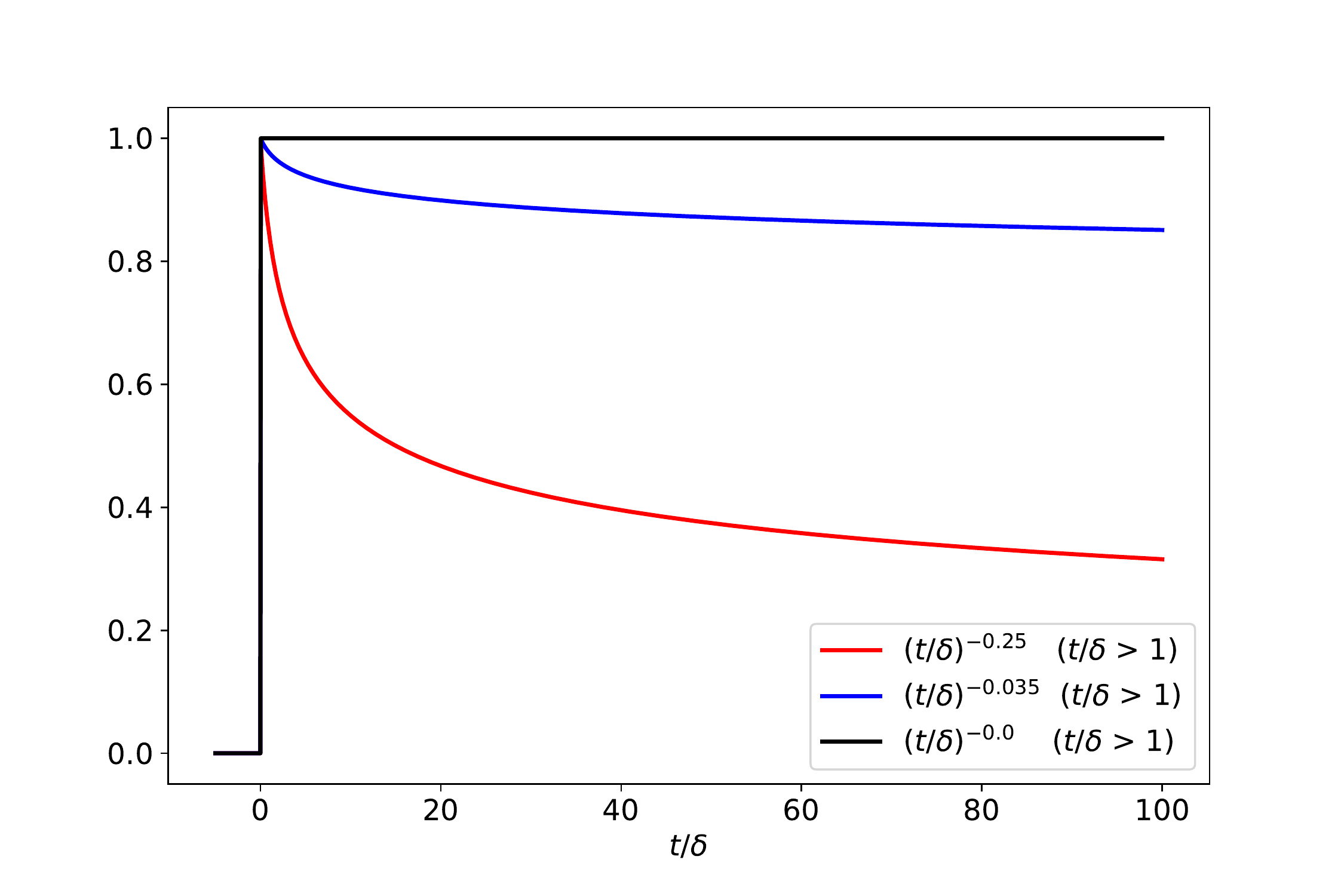}
\caption{Impulse responses for long-memory shot-noise processes. Exponent zero (black) yields a memoryless random walk. Small negative exponents yield nearly a random walk, but with a partial tendency to return to previous values.}
\label{fig0}
\end{figure}

Evidently, if the impulse $f(.)$ is a unit step function at argument zero, the process $p(t)$
is a random walk with steps of normally distributed amplitude $\sigma$ occuring randomly at a mean rate $\lambda$. Then, likewise obvious, is the implied variogram,
\begin{equation}
    V(\uptau) = \lambda \sigma^2 \,\uptau
\end{equation}
As is well known, a Gaussian random walk (Wiener or Brownian motion process) is obtained by taking the simultaneous limits $\lambda\rightarrow\infty$, $\sigma\rightarrow 0$, with $\lambda \sigma^2$ held constant. \cite{Rice77}.

A long-memory shot-noise process that is not a random walk is obtained by replacing the unit step function by something else \cite{Petropulu,Beran}, for example, 
\begin{equation}
f(t) = \begin{cases}
0 & -\infty < t < \delta \\
\left(t/\delta\right)^{(1-\epsilon)} & \delta \le t
\end{cases}
\end{equation}
where $\delta > 0$ is to be thought of as a small value (the interval $0 < t < \delta$ being implicitly reserved for the rise of the impulse) and $-\tfrac{1}{2} < \epsilon < \tfrac{1}{2}$.
Figure \ref{fig0} shows examples of such power-law impulses.
It is conventional
to term $H = \tfrac{1}{2} - \epsilon$ the Hurst exponent \cite{Hurst}. A process with $H > \tfrac{1}{2}$ (or $\epsilon < 0$) is termed persistent. One with $H < \tfrac{1}{2}$ (or $\epsilon > 0$) is termed anti-persistent or mean-reverting. Both are long-memory processes. Only the case $H=\tfrac{1}{2}$ ($\epsilon = 0$) is memoryless, a random walk.

The variogram of a Hurst power-law shot-noise process can be shown to be \cite{Baillie,Qian}
\begin{equation}
    V(\uptau) \propto \uptau^{2H} = \uptau^{1-2\epsilon}
\end{equation}
so a model for the NYSE data in Figure \ref{fig2} or equation \eqref{eq2} is a Hurst power-law
process with $\epsilon \approx 0.035$.

In the Gaussian limit of many overlapping impulses, the Hurst power-law process is known as fractional Brownian motion \cite{Baillie,Lowen,MVN,Qian}.
The simulation of a long time series of fractional Brownian motion is straightforward: Fast Fourier Transform methods are used to convolve a series of i.i.d. normals ($s_i$'s in equation \eqref{eqshot}) with an impulse $f(.)$ as in Figure \ref{fig0}. On a desktop machine with a single GPU we can simulate 3600 years of hourly prices in a few wall-clock seconds.

\subsection{Why Use Variogram Instead of Autocorrelation?}

Variogram $V(\uptau) = \E_t[(p_{t+\uptau} - p_t)^2]$ and autocorrelation $\E_t[r_{t+\uptau} r_t]$ contain equivalent two-point statistical information. For a Gaussian process, either characterizes a process completely. In empirical studies of asset returns, the autocorrelation has been more frequently studied (e.g., \cite{Baillie,fama1986,fama1988permanent,Lo,boudoukh1994}) than the variogram or its twin the variance ratio test (e.g. \cite{lo1988,poterba1988}). 

The Hurst power-law process is a convenient platform for us to explore the minority view. Suppose we have $V(\uptau) = \uptau^{1-2\epsilon}$ with $\uptau > 1$ and we want to distinguish it from the random-walk null hypothesis $V(\uptau) = \uptau^{1}$. Then the difference, for small $\epsilon > 0$, is a signal of positive magnitude
\begin{equation}
    \Delta V = \uptau - \uptau^{1-2\epsilon} = \uptau (1 - \uptau^{-2\epsilon})
    \approx 2 \epsilon \,\uptau\log(\uptau)
\end{equation}
We want to detect this signal in the presence of the noise that is the sample variance of the the full $V(\uptau)$. This ``measurement variance of the variance" for any Gaussian process whose variance $\sigma^2$ is sampled $N$ times, is $2\sigma^4/N$ \cite{NR}, so
\begin{equation}
    \Var[V(\uptau)] = \frac{2}{N} V(\uptau)^2 = \frac{2}{N}\uptau^{2-2\epsilon}
\end{equation}
implying a signal-to-noise ratio
\begin{equation}
    S/N \approx \sqrt{2N}\,\epsilon\,\log(\uptau)
\label{eqsn1}
\end{equation}

The autocorrelation (now the signal) of the same Hurst power-law process is (\cite{Qian}, p.~26)
\begin{equation}
    \Corr(\uptau) = \E_t[r_{t+\uptau} r_t] = -2\epsilon(1-2\epsilon)\uptau^{-1-2\epsilon}
\end{equation}
which decays rapidly with increasing $\uptau$.
For small $\epsilon$, i.e., close to random walk, we can approximately calculate the noise variance as if $r_{t+\uptau}$ and $r_t$ were independent, so
\begin{equation}
    \Var[\Corr(\uptau)]  = \Var\left[\frac{1}{N}\sum^N r_{t+\uptau} r_t  \right]
    = \frac{1}{N^2} \sum^N \Var[r_{t+\uptau} r_t] = \frac{1}{N^2} \sum^N [\Var(r_t)]^2
    = \frac{1}{N}
\end{equation}
since the $r$'s are independent normals with unit variance. Now the signal-to-noise ratio is, to leading order in $\epsilon$
\begin{equation}
    S/N \approx 2\sqrt{N}\,\epsilon\,\uptau^{-1-2\epsilon}
\label{eqsn2}
\end{equation}
Comparing equations \eqref{eqsn1} and \eqref{eqsn2}, one sees that for the same values $\epsilon$ and $N$, the variogram measurement becomes logarithmically more statistically significant as $\uptau$ increases, while the autocorrelation becomes less significant, at least inversely with $\uptau$. That the significant part of the autocorrelation signal is concentrated at the shortest times is problematic also because that is where the data is most subject to artifacts (bid-ask bounce, spoofing, or layering, for example), while the variogram's use of price movements over longer times should be more reliable.

Depending on exactly the way the data is gathered, one might argue that there are only $N/\uptau$ independent samples of $V(\uptau)$, while there are fully $N$ samples for the autocorrelation. In that case, the advantage of the variogram is only $\sim\sqrt{\uptau}$ instead of $\sim\uptau$, but it is still an advantage.

\subsection{Arbitrage Trading Strategy for Fractional Brownian Motion}
\label{sechurstarb}

One expects a fractional Brownian process with $\epsilon \ne 0$ to be arbitrageable, because it is not memoryless. Since the observed NYSE exponent $\epsilon$ is positive, our model is mean-reverting. Thus, we expect to make a profit by betting against its returns, that is, going long after a negative return, short after a positive one.

Our numerical arbitrage test first generates a discrete fractional Brownian vector of length 31,536,000 with Hurst exponent $\epsilon = -0.035$. The values are interpreted as 3600 years
of hourly samples, 8760 (transaction time) hours per year. Each year separately is linearly detrended to begin and end at the logarithmic value zero. Next, all values are multiplied by a factor that gives their standard deviation (across all hours and years) the value $0.15$, this to approximate the annualized volatility of log prices in the NYSE. The scaled values are finally exponentiated to produce hourly prices $p_{y,h}$, with $y$ indexing the year, $h$ the hour. Since each year begins and ends with a price of one (``\$1.00"), a simple buy-and-hold trading strategy will yield exactly zero return in each year. A useful normalizing value below is the r.m.s.~1-hour return across the whole sample,
\begin{equation}
    r_\text{rms} \equiv \E_y\left[ \left( \E_h\left[ \log(p_{y,h+1}/p_{y,h})^2 \right]\right)^{1/2}\right]
\end{equation}
where expectations $\E$ are estimimated by sample means.

We next define a trading strategy. Since the actual NYSE data consists of interval averaged prices (see discussion in  \S\ref{sec1p2}, \S\ref{sec1p22}, and \S{}S1), we interpret the simulated prices in the same way. We must take care to be fully causal, implying that no use can be made of an interval's average price until after the end of the averaging interval. We adopt the following trading strategy:

\vspace{-\parskip}
\vspace{4pt}
For every year $y$ and hour $h$,
\vspace{-4pt}
\begin{itemize}
    \setlength\itemsep{0pt}
    \item Using $h$ and $h+1$ prices, calculate normalized returns: $\hat r_{y,h} =\log(p_{y,h+1}/p_{y,h})/r_\text{rms}$
    \item Using $h+1$ prices, calculate the number of shares to buy (positive) or sell short (negative):
    $q_{y,h} = -\hat r_{y,h}/p_{y,h+1}$. The minus sign embodies the bet on a mean-returning process.
    \item Buy (sell) these shares during hour $h+2$, realizing that hour's average price $p_{y,h+2}$
    \item Sell (buy) the same shares during hour $h+3$, realizing that hour's average price $p_{y,h+3}$
    \item Record a profit or loss $P_{y,h} = q_{y,h} (p_{y,h+3}-p_{y,h+2})$
\end{itemize}
Note that the same (on average) capital is recycled in each hour and is at risk for
on average one hour between buying during $h+2$ and selling during $h+3$. We will be long and short about equally often, so it is a matter of convention how to define a denominator for the purpose of calculating annual returns. If we use the mean of absolute values $|\hat r_{y,h}|$,
counting long and short capital as equally at risk, then each year's net return is
\begin{equation}
    P_y = 8760 \sum_h P_{y,h} \bigg/ \sum_h |\hat r_{y,h}|
\end{equation}
This is an uncompounded return because the hourly stake is not increased (decreased) with profit (loss).

\begin{figure}[ht]
\centering
\includegraphics[width=16cm]{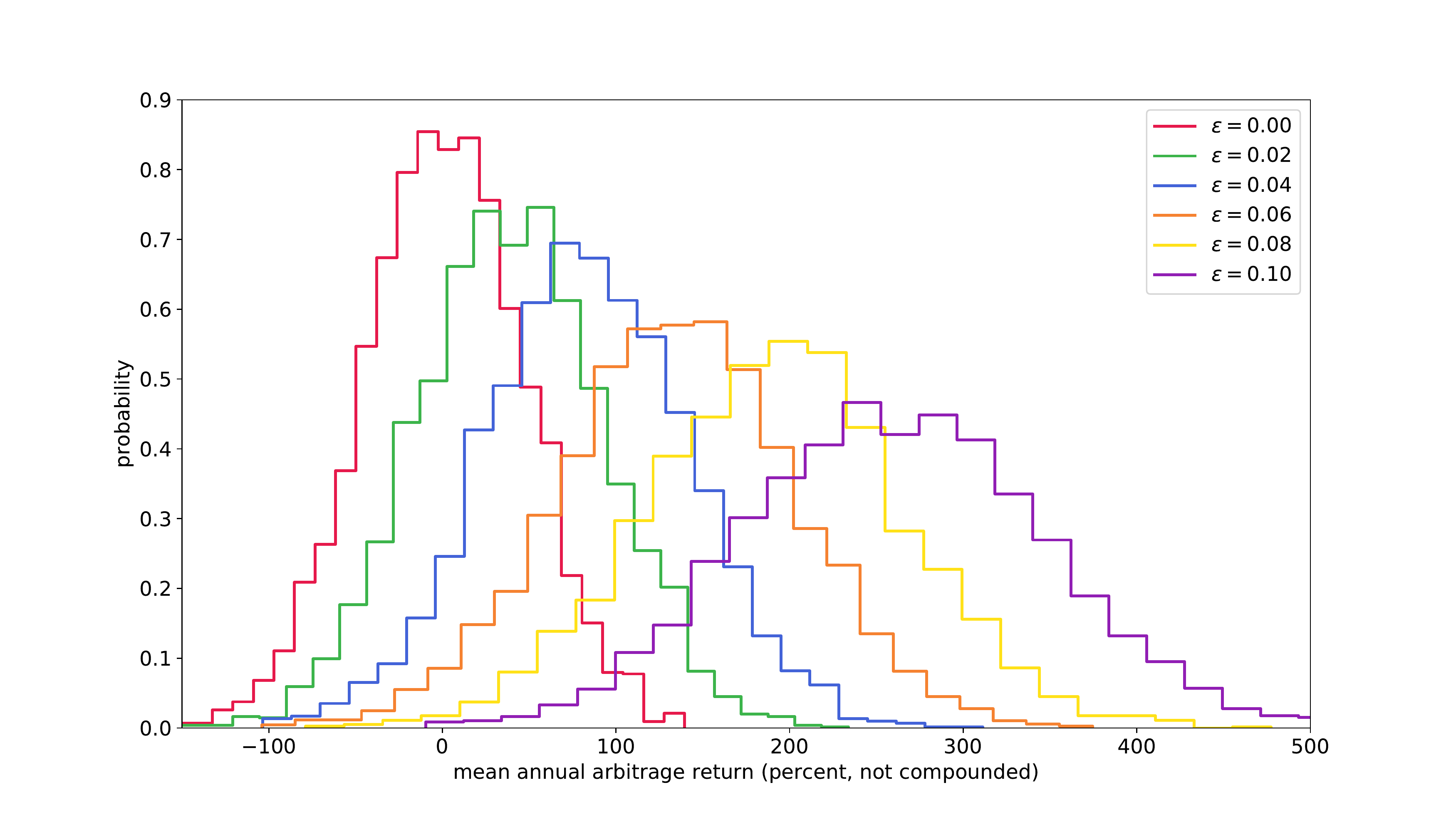}
\caption{Yearly net returns for a simple long-short arbitrage strategy applied to simulations of fractional Brownian motion over 3600 1-year periods with hourly data in each. The Hurst exponent for the six $\epsilon$ cases shown is $H = 0.5 - \epsilon$. The case $\epsilon = 0$ is a true random walk for which arbitrage should not (on average) show a net profit, as is seen.}
\label{fighurst1}
\end{figure}

\begin{figure}[ht]
\centering
\includegraphics[width=16cm]{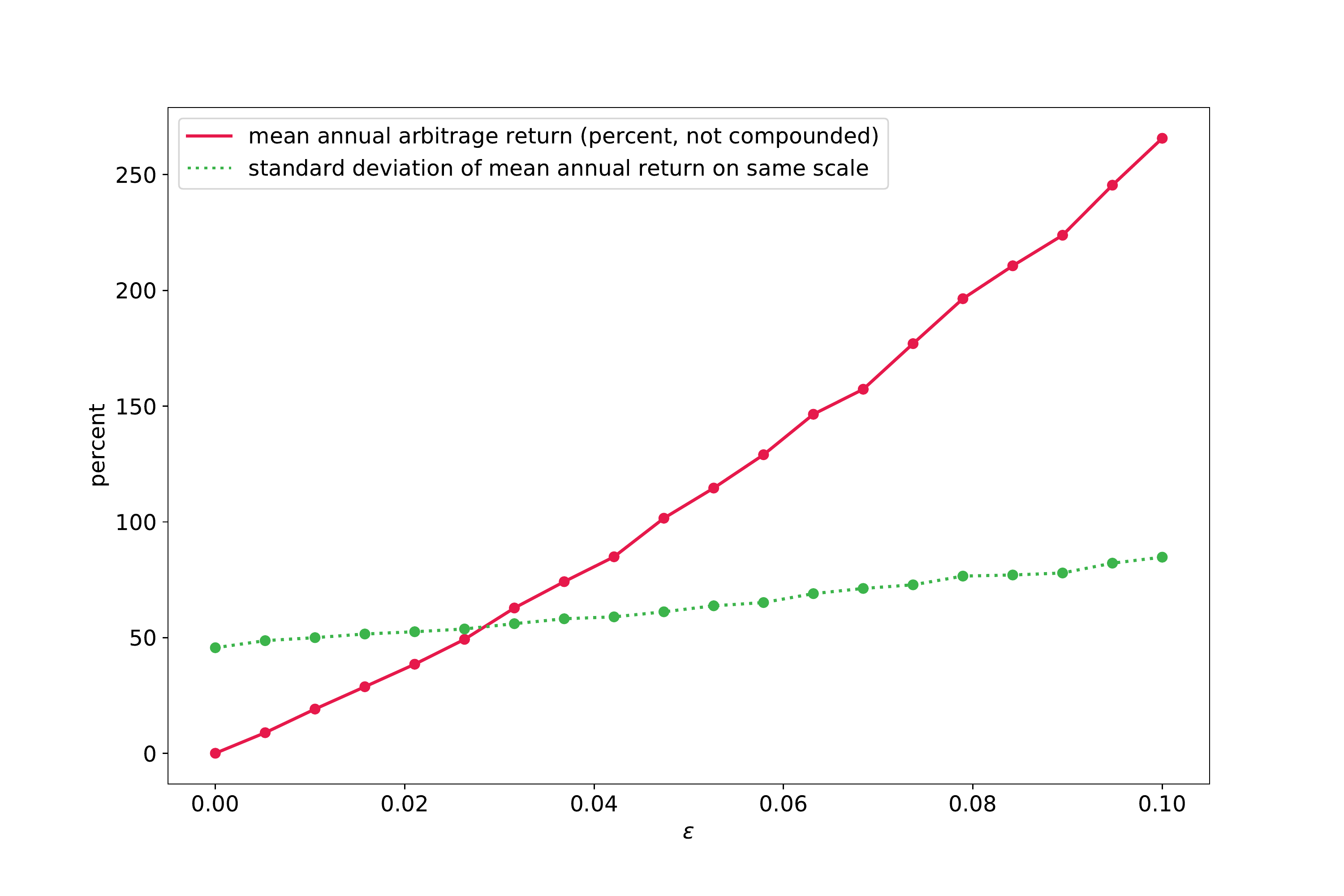}
\caption{Means and standard deviations of annual net arbitrage returns like those shown in Figure \ref{fighurst1}, but now plotted as functions of $\epsilon$. One sees that at $\epsilon \approx 0.03$, an arbitrage signal is detectable at about one standard deviation (1-$\sigma$).}
\label{fighurst2}
\end{figure}

Figure \ref{fighurst1} shows the histograms of yearly net returns $P_y$ for the 3600 simulated years, for values of $\epsilon$ between 0 and $0.10$ (Hurst exponents between $0.5$ and $0.4$). For the case $\epsilon = 0$, one sees returns roughly symmetrically around zero, as one expects for this memoryless random walk. For larger values of $\epsilon$, the adopted trading strategy shows significant positive returns.

Figure \ref{fighurst2} shows results from simulations with a larger number of $\epsilon$ values and plots their mean return and the standard deviation (over 3600 simulated years) of that mean. That the mean return for the random walk case $\epsilon = 0$ is close to zero (value $\approx 0.003$) is a useful check, since hourly long and short trades were simulated as for other values of $\epsilon$. One sees that a detection at one standard deviation with one-year's data becomes possible when $\epsilon$ exceeds about $0.03$

The conclusion of this section is that a model long-memory power-law process with $\epsilon$ as small as $\sim 0.01$ can be recognized by a simple arbitrage trading strategy over one (or a few) years of hourly price data. In particular, that true process is thus easily distinguished from a memoryless process whose apparent nonzero $\epsilon$ is due to a poor choice of transaction time or any other unmodeled cause, since no profitable arbitrage should be possible in that case.

\FloatBarrier

\section{Arbitrage Strategy Applied to the NYSE Data}
\label{secarbit}

We are now in a position to apply in backtesting of real data (no longer simulation!) something like \S\ref{sechurstarb}'s trading strategy. In simulation, we looked at a single (imaginary) stock over some thousands of homogeneous years. For the actual data, we have about a thousand stocks, each over five inhomogeneous years (e.g., COVID-19 occurred). Our trading algorithms will therefore be different in detail, though similar in spirit.

\begin{figure}[ht]
\centering
\includegraphics[width=16cm]{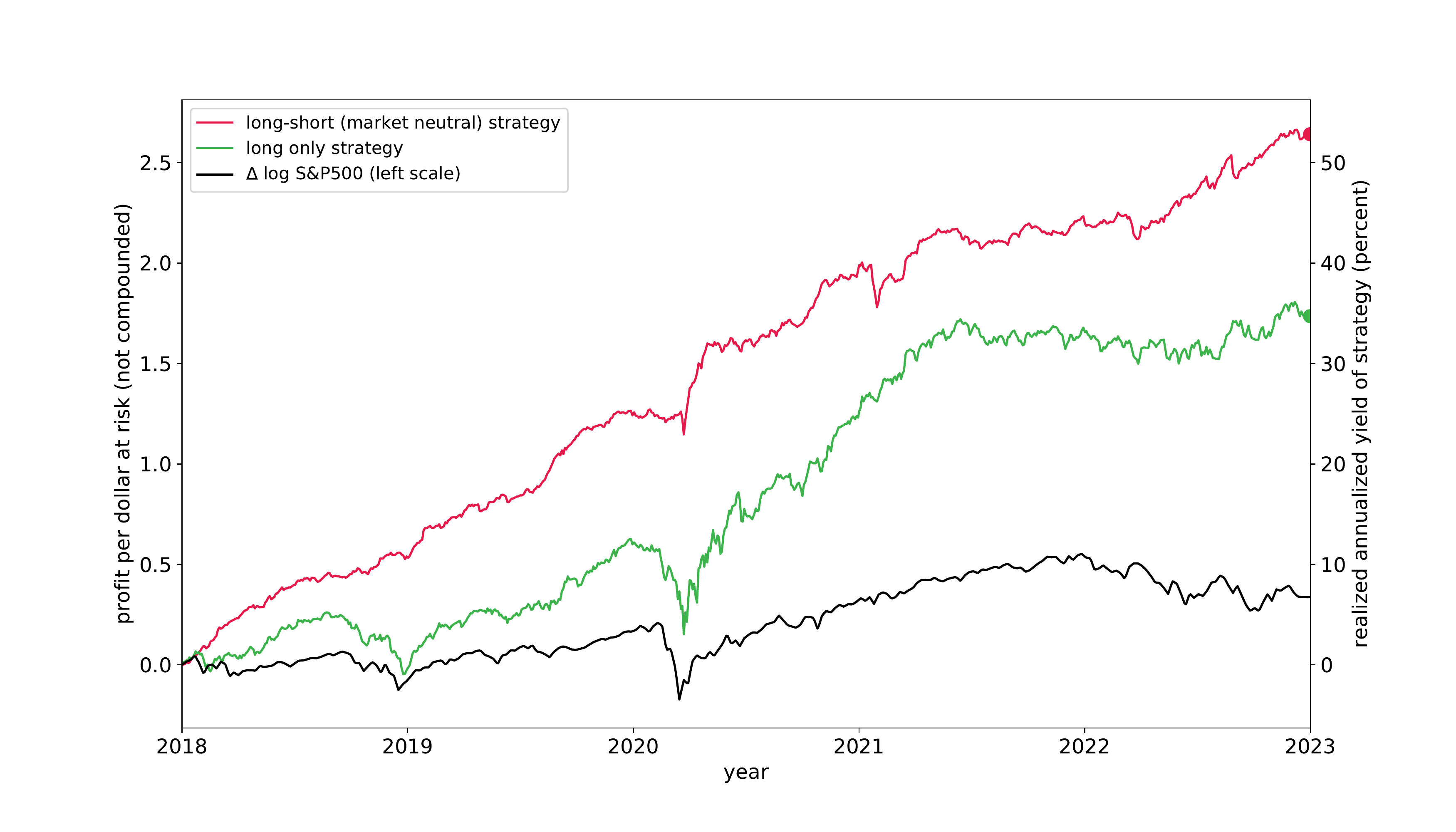}
\caption{Backtesting historical data, curves show the growth of a unit stake (uncompounded) under an arbitrage trading strategy that posits a long-memory power-law model with mean-regressing behavior. Also shown is growth of the S\&P 500 index on the same scale. The right-hand scale shows the implied annualized yield of the end-point of each curve.}
\label{fig3}
\end{figure}

\vspace{-\parskip}
\vspace{4pt}
We proceed as follows: For each successive 1-hour (\DTW) interval (indexed $h$) in years 2018--2022, and for the 1,091 NYSE stocks that traded during all five years for which we have minute-resolution data,
\vspace{-4pt}
\begin{itemize}
    \item Eliminate from consideration stocks that traded in fewer than half of all \DTW~hours in one or more years. (Such stocks produce too much missing data in our following use of consecutive hourly prices.) This cut on the data results in 866 stocks remaining, now indexed by $k$.
    \item For each stock $k$, observe its average trading price in hours $h$ and $h+1$ and compute a return $r_{k,h}=\log(p_{k,h+1}/p_{k,h})$.
    \item If fewer than 100 stocks have positive returns, or fewer than 100 have negative returns, take no action in hour $h$. (This to avoid unusual trading hours with massive correlated movements.)
    \item Otherwise, divide a fixed stake among stocks with negative $r_{k,h}$'s proportional to $|r_{k,h}|$ and purchase those stocks during hour $h+2$, as close as possible to uniformly in \DTW time. We make the assumption that an average price very close to the average trading price of the hour can be obtained. Similarly go short in the same total amount on stocks with positive $r_{h,k}$'s, in proportion to their $|r_{h,k}|$.
    \item Uniformly in \DTW during hour $h+3$ close all positions acquired in hour $h+2$, so that the mean holding time is 1 hr.
\end{itemize}
The above prescription is applied every hour, so that, as in the Hurst simulation, we are at various stages of four different hours at any given time. However, our outstanding long stake (and equal short risk) is constant as one hour's worth; specifically, we are again not compounding returns (or losses). Also worth emphasizing again is that our strategy is completely causal: In any hour, we take actions based only on previous hours, not the present one.

How do we do? Figure \ref{fig3} shows the result for the above long-short strategy, and also for a strategy where only the long trades (on stocks with a negative return in the previous hour) are made. Both strategies do significantly better than the market (S\&P 500 as a surrogate) during almost all intervals, and very much better cumulatively over five years. In particular, the long-short strategy returns 53\% (annualized), the long-only strategy 35\%. The COVID-19 crisis in March-April, 2020 produces a brief downturn in the long-only strategy, but is barely visible (and mostly with a positive effect) in the long-short strategy.

For many reasons, we should not expect exact agreement between these annual arbitrage returns and the Hurst simulations. The trading strategies are, of necessity, not identical. The value $0.15$ used in the simulations for the market's annualized volatility is indicative, but not exact, especially not for any particular stock. The actual market's cross-correlational structure can (as we will see in \S\ref{sec6} below) play a significant role. Still, it is encouraging that the actual data, with measured $\epsilon\approx 0.035$, and the simulations (Figure \ref{fighurst2} at that $\epsilon$ value) are in the same ballpark with $\sim 60$\% annual return. 

We conclude that the long-memory power-law model (or fractional Brownian motion model) with $\epsilon\approx 0.035$ (or Hurst exponent $H=0.465$) is quite a plausible approximation. In \S\ref{sec7} we will show further evidence, based on cross-correlational structure.

The above arbitrage opportunity may or may not be realizable in practice. Since a given dollar turns over $\sim 8000$ times per year, round-trip transaction costs $\lesssim 10^{-4}$ would be required, and the ability to trade $\sim 10^3$ each hour. This might be achievable with a large, market-making investment bank as a continuous counter-party, but not seemingly otherwise. Also, we assume that our purchases and short sales do not move the market, and that they can be accomplished during a trading hour at the average price of that hour, both assumptions open to reasonable question.

\FloatBarrier

\section{Cross-Correlation Structure}
\label{sec6}

Up to now, we have dealt with stocks one a a time, with a time series of prices for each. Now, we will look at the cross-correlations of multiple stocks. We define the covariance of two stocks $A$ and $B$ on timescale $\uptau$ by
\begin{equation}
    C_{AB}(\uptau) \equiv \E[r_A(\uptau) r_B(\uptau)] =
    \E_t \left[ (P_{t+\uptau}^{(A)} - P_t^{(A)})(P_{t+\uptau}^{(B)} - P_t^{(B)}) \right]
\label{eq15}
\end{equation}
where the $P$'s denote $\uptau$-averaged stock prices, as described in \S\ref{sec1p22},
and $\mathbf{C}$ is the covariance matrix.
We will use the terms cross-correlation and correlation interchangeably as meaning
\begin{equation}
    \rho_{AB}(\uptau) \equiv \frac{C_{AB}(\uptau)}{\sqrt{C_{AA}(\uptau)C_{BB}(\uptau)}}
\label{eq16}
\end{equation}
(We continue to use the term autocorrelation to mean correlation in time of a single stock.)
Because not every stock trades every minute, we need a method for estimating the
expectations $\E$ from possibly asynchronous trading data derived from one-minute candles. For this we use the formalism developed in \cite{press2023}.

Also from \cite{press2023} we adopt the shorthand notation
$\Nor$ as denoting a draw from the normal distribution $N(0,1)$ and adopt the convention that different subscripts $\{\Nor_X,\Nor_Y\}$ represent independent draws, identical subscripts $\{\Nor_Z,\Nor_Z\}$ denote the same draw (that is, have the same numerical value).

With these conventions, the Gaussian model return $r(\uptau)$ of a single uncorrelated stock $A$ (the model of \ref{sec2} above) can be written
\begin{equation}
    r_A(\uptau) \sim \Vh(\uptau) \Nor_A
\end{equation}
where $\sim$ is read ``is drawn from", while the correlated returns of two stocks $\{A,B\}$can be written
\begin{equation}
\begin{aligned}
    r_A(\uptau) &\sim \Vh_A(\uptau)\,\left( \sqrt{\rho_{AB}} \Nor_{C} + \sqrt{1-\rho_{AB}} \,\Nor_A  \right) \\
    r_B(\uptau) &\sim \Vh_B(\uptau)\,\left( \sqrt{\rho_{AB}} \Nor_{C} + \sqrt{1-\rho_{AB}} \,\Nor_B  \right)
\end{aligned}
\label{eq18}
\end{equation}
Here $\sqrt{\rho}$ is shorthand for $\pm\sqrt{|\rho|}$ with a minus sign only for negative correlations and on one stock, left to the reader to fill in appropriately. While equation \eqref{eq18} is written as a single factor model, it is in fact general for a Gaussian process, because the latter is completely defined  by its two-point correlations. The advantage of this notation is to compactly specify both autocorrelation in time $V(\uptau)$ and cross-correlation across stocks ($\rho_{AB}$).

\subsection{NYSE Cross-Correlations at 1-Hour Resolution}

With the methods described, we can readily compute $\rho_{AB}(\uptau = 1 \text{ hour})$ across all (\DTW) hours in the five-year period 2018--2022 for all pairs $A,B$ of the 866 NYSE stocks with one-minute candle data in all years. The resulting $866\times 866$ symmetric matrix is understandably difficult to visualize, but its related metric correlation distance $d_{AB} = \sqrt{1 - \rho_{AB}}$ implies an exact embedding, positioning the individual stocks in 865-dimensional space. (Analogously, a triangle of three correlations can always be embedded in two dimensions.) This 865-dimensional space can then be projected into two dimensions in various ways. Informative qualitative patterns emerge, for example, correlational relations among different industries. Supplementary Information \S\ref{secatlas} provides an ``Atlas" of figures generated in this way, but the idea of projection is not further used here.

\begin{figure}[ht]
\centering
\includegraphics[width=14cm]{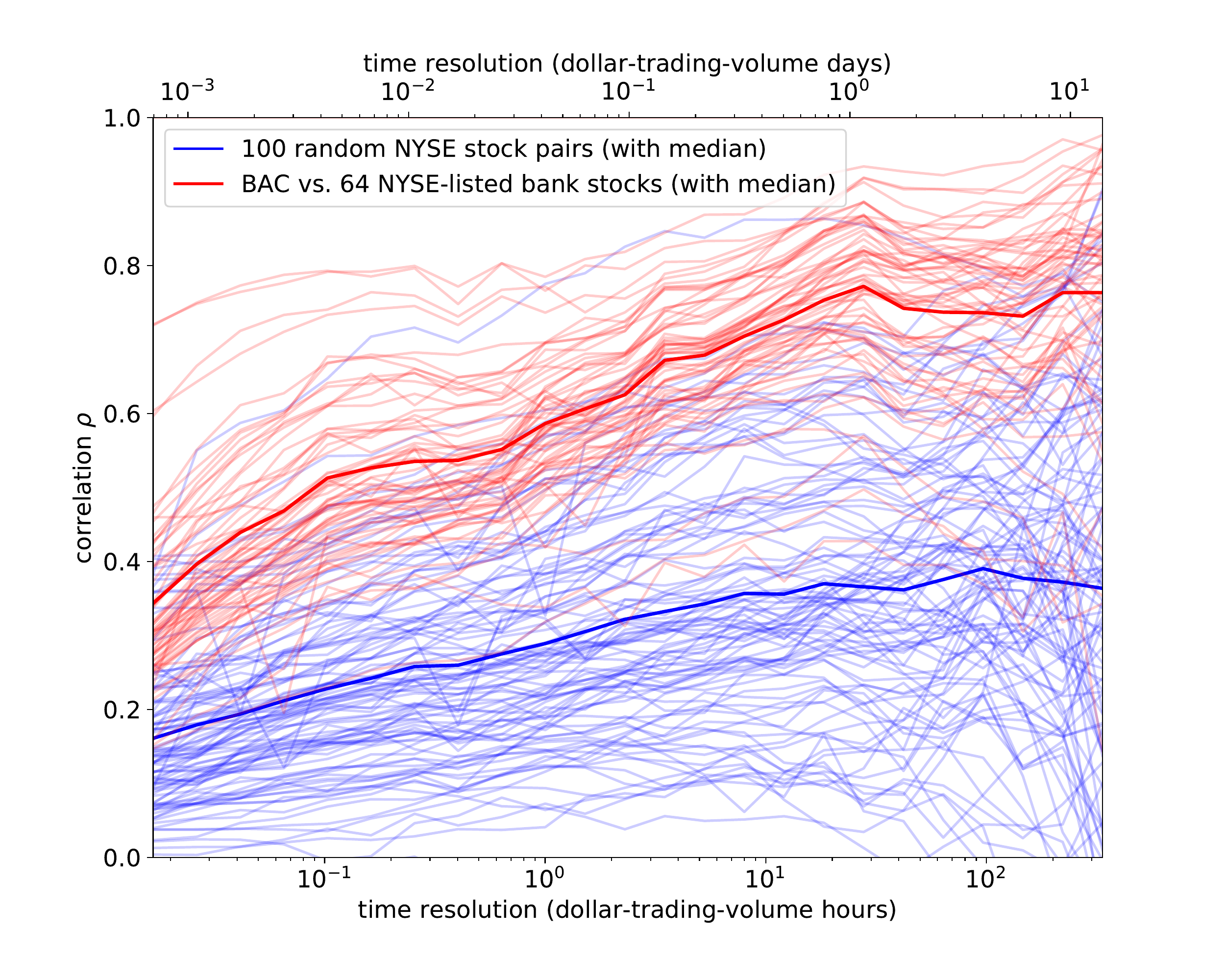}
\caption{Pairwise correlations $\rho$ for 100 randomly chosen pairs of NYSE stocks (red) as a function of time resolution $\uptau$, and correlations of 64 NYSE-listed bank stocks (blue) to Bank of America. Medians are shown as darker lines. The input data is year 2018 one-minute candles. The rising trend with $\uptau$ of essentially all correlations is to be noted.}
\label{figrandom100}
\end{figure}

\subsection{NYSE Cross-Correlations as Functions of Resolution $\uptau$}
\label{sec62}

As initial orientation, Figure \ref{figrandom100} shows, as a function of time resolution inverval $\uptau$, the pairwise correlations $\rho$ for 100 randomly chosen pairs of NYSE stocks, and also the median correlation. Similarly shown are the correlations of 64 NYSE-listed bank stocks with Bank of America (ticker BAC). Unsurprisingly the latter correlations, between banks, $\sim 0.7$, are larger than the former, between random firms, $\sim 0.3$. Noteworthy is that both sets of correlations increase, roughly linearly in $\log\uptau$, between $\sim 2$ min and $\sim 24$ hr, a range of almost three orders of magnitude.

\FloatBarrier

\begin{figure}[ht]
\centering
\includegraphics[width=14cm]{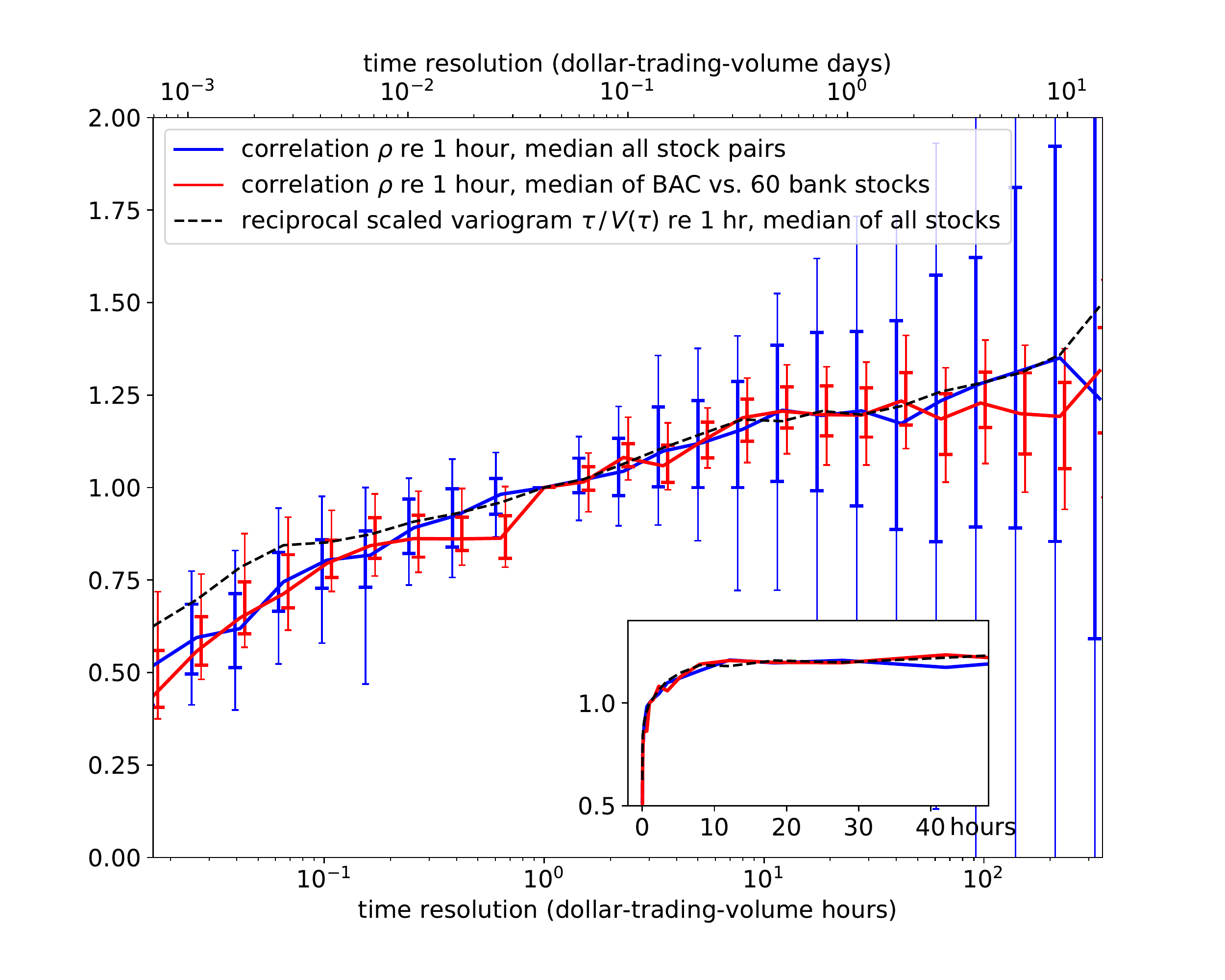}
\caption{Pairwise correlations as a function of resolution interval $\uptau$, now normalized to the value 1 at $\uptau = 1$ hour. Percentile sticks show 10\%, 25\%, 50\%, 75\%, and 90\% percentile points for all pairs of stocks at each value $\uptau$. A surprisingly universal pattern of increase in $\rho$ is seen. The inset linear-linear plot of the median curves illustrates that most of the increase is in the shortest times. See text for explanation of dashed black curve.}
\label{figrandom10000}
\end{figure}

Figure \ref{figrandom10000} extends this observation across all pairs of NYSE stocks, but now normalizing the values $\rho(\uptau)$ to a value 1.0 at $\uptau=1$ hour. Instead of individual stock pairs, percentile sticks are shown, the same percentile points as Figure \ref{fig2} above. One sees a remarkable degree of universality in the pattern of increase of correlations $\rho$. However, the mere fact of increase is itself at first sight puzzling, since a simple correlation model, equation \eqref{eq18} immediately implies $\rho_{AB}(\uptau) = \text{constant}$, since the factors in $V(\uptau)$ cancel in equation \eqref{eq16}.

Instead, the data in Figure \ref{figrandom10000} call for a model in which the correlated and uncorrelated parts of stocks' returns have different variogram dependences on $\uptau$. The simplest such model is
\begin{equation}
\begin{aligned}
    r_A(\uptau) &\sim \Vh_{A}(\uptau)\,\left( \sqrt{\rho_{AB}} \Nor_{C} + \sqrt{1-\rho_{AB}} \,\Nor_{A_1} \right) + \Uh_{A}(\uptau)\,\Nor_{A_2} \\
    r_B(\uptau) &\sim \Vh_{B}(\uptau)\,\left( \sqrt{\rho_{AB}} \Nor_{C} + \sqrt{1-\rho_{AB}} \,\Nor_{B_1} \right) + \Uh_{B}(\uptau)\,\Nor_{B_2}
\end{aligned}
\label{eq19}
\end{equation}
where $V$ and $U$ are respectively the correlated an uncorrelated partial variograms.
From equation \eqref{eq19} one readily calculates the total variograms $V^\text{tot}(\uptau)$ and related quantities,
\begin{equation}
\begin{aligned}
    V^\text{tot}_A(\uptau) &= \E[r_A r_A] =  V_{A}(\uptau) + U_{A}(\uptau)\\
    V^\text{tot}_B(\uptau) &= \E[r_B r_B] =  V_{B}(\uptau) + U_{B}(\uptau)\\
    \E[r_A r_B] &= \Vh_{A}\,\Vh_{B}\,
    \rho_{AB} \\
    \Rightarrow \rho^\text{observed}_{AB}(\uptau)
    &= \left( \frac{V_{A}(\uptau)}{V^\text{tot}_{A}(\uptau)}\;\frac{V_{B}(\uptau)}{V^\text{tot}_{B}(\uptau)} \right)^{1/2} \;\rho_{AB}
\end{aligned}
\label{eq20}
\end{equation}

Let us now make some rash simplifying assumptions, and see whether they are compatible with the data. Assume that the correlated and uncorrelated variograms $V(\uptau)$ and $U(\uptau)$, with $V^\text{tot}(\uptau) = V(\uptau) + U(\uptau)$, have a universal functional form across all stocks, and are only scaled by a factor for each stock. For example, they might each be a Hurst power-law process or a random walk. Then by equation \eqref{eq20},
\begin{equation}
    \frac{\rho^\text{observed}_{AB}(\uptau)}{\rho_{AB}} = \frac{V(\uptau)}{V^\text{tot}(\uptau)}  = \frac{1}{1 + \frac{U(\uptau)}{V(\uptau)}}
\label{eq21}
\end{equation}
Next rashly assume that the correlated part of the variance is memoryless, $V(\uptau)\propto \uptau$, so that all of the arbitrageable signal seen in \S\ref{sec2} and exploited in \S\ref{secarbit} is due to the uncorrelated variance $U(\uptau)$. Then,
\begin{equation}
    \frac{V^\text{tot}(\uptau)}{\uptau} =  \frac{V(\uptau)}{\uptau}
    \left[1 + \frac{U(\uptau)}{V(\uptau)} \right]    \propto 1 + \frac{U(\uptau)}{V(\uptau)}
\label{eq22}
\end{equation}
Equations \eqref{eq21} and \eqref{eq22} make a prediction,
\begin{equation}
    \rho^\text{observed}_{AB}(\uptau) \propto \frac{\uptau}{V^\text{tot}(\uptau)}
\label{eq23}
\end{equation}
This is directly testable in the data, without any assumption about the functional form of $U(\uptau)$  (e.g., Hurst or fractional Brownian). Indeed, we already plotted $\uptau/V^\text{tot}(\uptau)$ (normalized at 1 hour) in Figure \ref{figrandom10000} as the dashed black line. It is virtually indistinguishable from the median $\rho$ (also so normalized), just as predicted by equation \eqref{eq23}

We conclude that both the deviations in variance from a random walk seen in Figure \ref{fig2} and the deviations from constant-in-time correlations seen in Figure \ref{figrandom10000} can be explained by a mean-reverting component to returns that is uncorrelated across stocks and distinct from a (close-to?) random-walk component embodying the cross-correlational structure. The roughly factor of two fall (rise) seen in Figure \ref{fig2} (Figure \ref{figrandom10000}), suggests that the uncorrelated and correlated variances are comparable in magnitude at timescales of order minutes, but that the uncorrelated part largely decays away by timescales of order hours. In \S\ref{secarbit} above we successfully (at least in backtesting) arbitraged the decaying component, one stock at a time. In the next section we attempt to improve the arbitrage performance by now using the full correlational structure.

\section{Leave-One-Out Predictions of Stock Returns}
\label{sec7}

The positive arbitrage returns in \S\ref{secarbit} derived from an algorithm for identifying, in a fixed hour $h$, stocks that were over- or under-valued in a predictive sense. The algorithm was simple: How did the price change between the two previous hours $h-2$ and $h-1$? The market's correlational structure implies another, somewhat orthogonal, algorithm for identifying possibly over- or under-valued stocks: Given a correlation structure, how compatible is a stock's return in a period with the returns of all other stocks in that same period? Specifically, how different is its return from a correlational prediction of its return.

This is prediction in a very limited sense, using some data at one time to predict other data at the same time. It is then an open question, to be answered experimentally, whether such an algorithm is predictive of the future. That is what this section attempts to elucidate.

\subsection{Equations for Leave-One-Out Prediction}

Suppose we want to predict a return $r_I$ as a linear combination of other returns $r_K$ (uppercase subscripts denoting different stocks),
\begin{equation}
    \widehat r_I = \sum_{K\ne I} B_{IK} \,r_K
\label{eq24}
\end{equation}
Minimizing the expected mean square error of the discrepancy for $B_{IK}$,
\begin{equation}
    \argmin_{\mathbf{B}}\E\left[  \left(\widehat  r_I - \sum_{K\ne I} B_{IK} \,r_K\right)^2  \right]
\label{eq25}
\end{equation}
yields after some algebra,
\begin{equation}
\begin{aligned}    
    B_{IK} &= \left[\left[  \E(r_K r_J)_{K,J\ne I} \right]\right]^{-1}
    \left[\left[  \E(r_J r_I)_{J\ne I} \right]\right]\\
    &= \sum_{J\ne I} \left[\bm{C}^{\ne I}\right]^{-1}_{KJ} C_{JI}\\
    &\equiv \sum_J A^{\ne I}_{KJ}C_{JI}
\end{aligned}
\label{eq26}
\end{equation}
Here double brackets indicate matrices, $\mathbf{C}$ is the correlation matrix
of equation \eqref{eq15}, and and $\bm{C}^{\ne I}$ means $\mathbf{C}$ with row and column $I$ removed. The matrix $A^{\ne I}_{KJ}$ is then most easily defined in words:

\vspace{-8pt}
\begin{itemize}
    \setlength\itemsep{-4pt}
    \item Delete row and column $I$ in the covariance matrix $\mathbf{C}$.
    \item Invert the resulting matrix.
    \item Insert a new row and column $I$ of all zeros to get $A^{\ne I}_{KJ}$
\end{itemize}
The formula for the inverse of a partitioned matrix can next be applied to give the simple result
(see, e.g., \cite{whuber}),
\begin{equation}
    A^{\ne I}_{KJ} = A_{KJ} - A_{JI}A_{KI}/A_{II}, \quad \text{where } \mathbf{A} \equiv \mathbf{C}^{-1}
\label{eq27}
\end{equation}
yielding after substitution into equation \eqref{eq26} and some algebra a particularly simple form for coefficients in the prediction equation \eqref{eq24},
\begin{equation}
    B_{IK} = -\left( \frac{A_{KI}}{A_{II}} - \delta_{KI}\right)
\label{eq28}
\end{equation}
The delta term merely zeros the diagonal elements, enforcing that a return is not used in predicting itself.
The manipulations in equations \eqref{eq27}--\eqref{eq28} are not just for elegance: They avoid having to separately invert a (say) $865\times 865$ matrix 866 times, for each value of $I$ as equation \eqref{eq26} seemingly indicates. Instead we can invert an $866\times 866$ matrix just once.

A figure of merit for the predictions is the mean (across stocks) fraction of variance explained (FVE) of hourly returns, equal to the square of the correlation coefficient between $r_{hI}$ and $\widehat  r_{hI}$. This can
be written as (see \S\ref{suppinfo13}),
\begin{equation}
    \text{FVE} = \left< \left[ 1 - \frac{1}{2}\frac{\left< (\widehat r_{hK} -  r_{hK})^2\right>_h}
    {\left< r_{hK}^2\right>_h} \right]^2 \right>_K
\label{eq29}
\end{equation}
where angle brackets denote sample averages over stocks $I$ and hours $h$ as indicated.
A perfect prediction would have $\text{FVE} = 1$.

\subsection{Leave-One-Out Results by Year}

Table 1 shows results where the covariance matrix $\mathbf{C}$ is estimated using one calendar
year's hourly data, matrix-inverted, and then used to predict each year's hourly returns on 866 stocks. 

\begin{table}[h]
\begin{center}
\begin{tabular}{p{5pt}c|ccccc}
\multicolumn{2}{}{} & \multicolumn{5}{c}{Year Predicted} \\
& & 2018  &  2019   & 2020    & 2021    & 2022 \\ \cline{2-7}
\multirow{5}{*}{\rotatebox[origin=c]{90}{Correlation Year}}
& 2018 & 63.1 & 40.6 & 48.3 & 43.7 & 52.0 \\
& 2019 & 44.1 & 60.8 & 49.0 & 44.2 & 52.2 \\
& 2020 & 39.7 & 37.0 & 71.7 & 42.2 & 50.0 \\
& 2021 & 43.4 & 40.5 & 50.3 & 63.9 & 54.7 \\
& 2022 & 42.7 & 39.7 & 49.2 & 46.2 & 70.8 \\ \cline{2-7}
& none & 37.7 & 34.3 & 45.0 & 36.2 & 43.8 \\
\end{tabular}
\caption{Fraction of Variance Explained (FVE), equation \eqref{eq29}, in percent. Each row uses hourly returns data from the
indicated year to construct an inverse correlation matrix. That matrix is used to predict hourly
leave-one-out returns of every stock in the year indicated by the column. Values are the
mean fraction of the variance of actual hourly returns of the stock explained. The bottom row shows results for a correlation matrix in which all pairs of stocks are given the same numerical correlation.}
\end{center}
\end{table}

On average about 66\%  of the hourly returns variance is explained using the same-year covariance---but this is subject to some over-fitting because the data is used twice. A more meaningful result is that, on average, about 45\% of the variance in different years is explained, and the value is about independent of which year. In other words, the correlational structure persists roughly unchanged over the five year period studied. Some years (e.g., 2022) are intrinsically more predictable than others (e.g., 2019). Interestingly, COVID-19 year 2020 has typical predictability, despite the market's extreme fluctuations.

As is well known, stock prices tend to move up or down in tandem. We might try predicting returns $r_{hK}$ by an unweighted average of all other returns, scaling by variance appropriately,
\begin{equation}
    \widehat r_{hK} = \sqrt{\Var(r_K)} \left< \frac{r_{hJ}}{\sqrt{\Var(r_J)}}\right>_{J\ne I}
\end{equation}
with angle brackets denoting sample mean. One sees, on average, about 39\% of variance explained, only fractionally less than with the full correlational structure. One can show that the correlation matrix corresponding to an unweighted average in equation \eqref{eq24} is one with equal positive off-diagonal terms, independent of their magnitude (if it is not too small).

\begin{table}[h]
\begin{center}
\begin{tabular}{p{5pt}c|ccccc}
\multicolumn{2}{}{} & \multicolumn{5}{c}{Year Predicted} \\
& & 2018  &  2019   & 2020    & 2021    & 2022 \\ \cline{2-7}
\multirow{5}{*}{\rotatebox[origin=c]{90}{Correlation Year}}
& 2018 & 59.3 & 46.1 & 53.7 & 49.1 & 56.6 \\
& 2019 & 49.3 & 56.6 & 54.2 & 49.2 & 56.6 \\
& 2020 & 48.1 & 45.5 & 65.8 & 49.9 & 56.8 \\
& 2021 & 48.9 & 46.1 & 55.4 & 60.2 & 58.8 \\
& 2022 & 49.1 & 46.2 & 55.2 & 51.5 & 65.7 \\ \cline{2-7}
& none & 37.7 & 34.3 & 45.0 & 36.2 & 43.8 \\
\end{tabular}
\caption{Same as Table 1, except that better predictions are obtained using gradient
backpropagation instead of direct covariance matrix inversion. On average, 51.3\% of the
variance of hourly returns is explained by any other year's correlation structure 
(61.5\% by the same year's).}
\end{center}
\end{table}

A technical issue is that, because all stocks tend to be positively correlated, with some pairs very highly correlated, the covariance matrix $\mathbf{C}$ is close to singular. Thus, the inversion implicit in Table 1 is quite noisy. With gradient backpropagation machinery standard in the training of large neural networks (NNs), we can calculate an improved matrix $A_{IK}$ by direct minimization of equation \eqref{eq25}. Doing this yields the improved results shown in Table 2. We inhibit overfitting by a standard neural-network technique: Each year's fit is gradient-minimized with only its own data, but the gradient search is terminated when the performance on other years is no longer decreasing.

The average off-diagonal value in Table 2, that is, the fraction of variance explained using a different year's correlation matrix, is 51.3\%, significantly better than the 45.4\% seen in Table 1. The average of the diagonals, predicting a year by its own correlation, is 61.5\%, lower than Table 1's 66.1\% because the overfitting is now less.

\begin{figure}[ht]
\centering
\includegraphics[width=16cm]{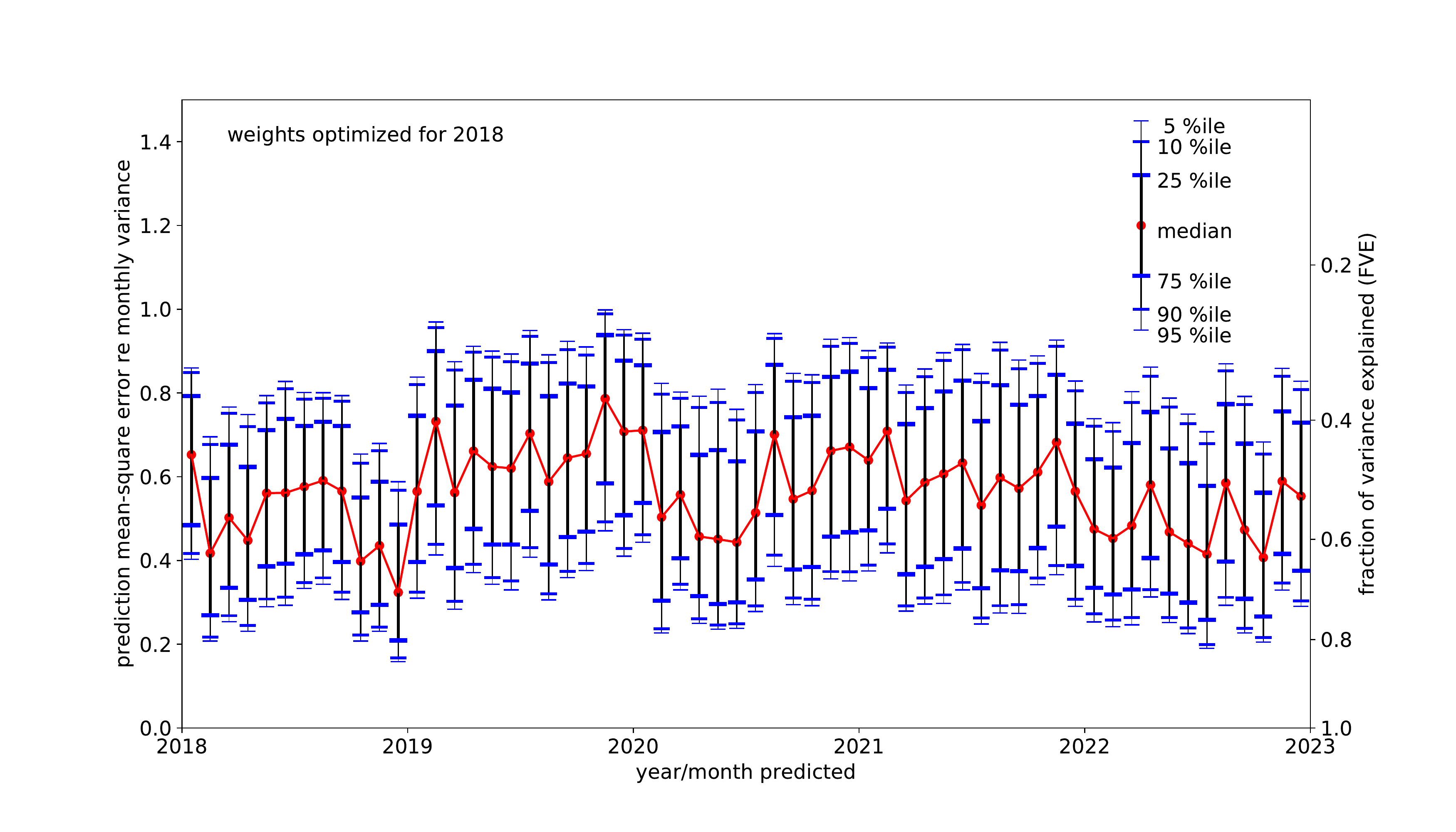}
\caption{Monthly fractional mean-square error (FMSE, left scale) and fraction of
variance explained (FVE, right scale) for correlational predictions of 1-hour returns of NYSE stocks. A single year (2018) is used to estimate the inverse correlation matrix, which is then applied across all years. The percentile sticks show the range of prediction accuracies over all 866 stocks.}
\label{figpredbymo2018}
\end{figure}

\subsection{Leave-One-Out Monthly Performance}

The percent values given in Tables 1 and 2 are fractions of the full year's variance explained (FVEs). But because a stock's (or the market's) volatility can vary significantly over the course of a year, it is also interesting to disaggregate the results of Table 2 by months and to show the fraction of each month's individual variance explained by correlational prediction. Figure \ref{figpredbymo2018} shows this and also the equivalent metric of fractional mean-square error, averaged over stocks,

\begin{equation}
    \text{FMSE} = \left< \frac{\left< (\widehat r_{hK} -  r_{hK})^2\right>_h}
    {\left< r_{hK}^2\right>_h} \right>_K
\label{eq31}
\end{equation}
In the Figure, the single year 2018 is used to estimate the inverse correlation matrix, which is then applied across all years. Percentile sticks show the ranges of predictabilities over all the stocks.

\begin{figure}[ht]
\centering
\includegraphics[width=16cm]{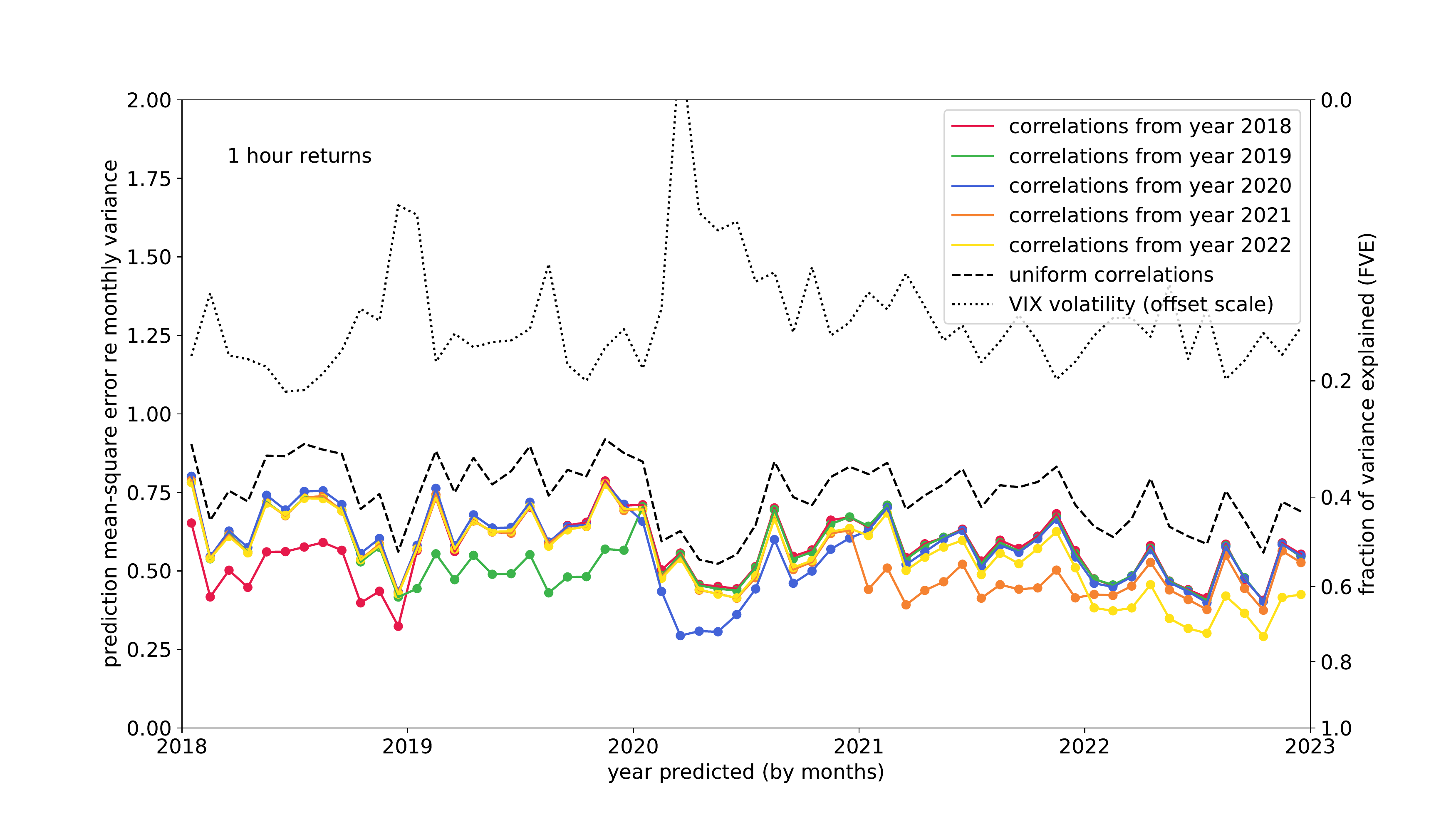}
\caption{Like Figure \ref{figpredbymo2018}, but using different single years (indicated by color) to estimate the inverse correlation matrix, which is then applied across all years. One sees that each year (color) does best in predicting itself, due to some overfitting, but that otherwise the correlations of any year are about equally good at predicting across all years shown. Also shown (on an arbitrary offset scale) is the VIX volatility measure. Predictions tend to be best when volatility is high.}
\label{figpredbymo}
\end{figure}

Figure \ref{figpredbymo} shows the results of a similar analysis, showing only the median across all stocks, but now using different single years (indicated by color) to to estimate the inverse correlation matrix.
One sees that, due to overfitting in using a year's correlation data to predict that same year, each year does best in predicting itself, but that otherwise the correlations of any year are about equally good at predicting across all years shown, confirming the general behavior seen in Tables 1 and 2.

Also shown in Figure \ref{figpredbymo} is the VIX volatility measure. Predictions tend to be best when volatility is high, and were especially good (measured by FVE or FMSE) in the COVID-19 period of high volatility. The figure also confirms (dotted curve) that the full correlations do somewhat, but not hugely, better than a simple unweighted average that corresponds to assuming an identical correlation among all stocks.

\FloatBarrier
\section{Arbitrage Strategy Using Cross-Correlations}
\label{sec8}
As was done in \S\ref{secarbit}, we can devise an arbitrage strategy, but now based on the same-time predictions of the correlation structure instead of on the mean-reversion memory of individual stocks. Now, we bet against any discrepancy between a stock's actual hourly return and the same-hour return correlational prediction of equations \eqref{eq24} and \eqref{eq28}.

In particular, we adopted this strategy: We computed prediction coefficients using 2018 hourly data and equation \eqref{eq28}. Then, for each successive 1-hour (\DTW{}) interval (indexed $h$) in years 2019-2022, and for the same 866 NYSE stocks (indexed $k$) as in \S\ref{secarbit},
\begin{itemize}
    \item For each stock $k$, observe its average trading price in hours $h$ and $h+1$ and compute a return $r_{k,h}=\log(p_{k,h+1}/p_{k,h})$.
    \item For each stock $k$, use the prediction coefficients to predict a return $\widehat r_{k,h}$ and from this a discrepancy $\Delta_{k,h} = \widehat r_{k,h} - r_{k,h}$
    \item Identify the 5\% of stocks with the largest (generally most positive) $\Delta_{k,h}$, and the 5\% with the smallest (generally most negative) discrepancy.
    \item Divide a fixed stake (i.e., investment) equally among the first list and purchase them uniformly during hour $h+2+S$, where $S \ge 0$ is a ``staleness" parameter. (In \S\ref{secarbit} we did only the case $S=1$.) Similarly go short on the second list in the total amount of the same stake.
    \item Uniformly during hour $h+3+S$ close all positions acquired in hour $h+2+S$ so that the mean holding time is 1 hr.
\end{itemize}

This strategy is not completely independent of the strategy used in \S\ref{secarbit}. There, a stock's discrepancy was with respect to a baseline expectation of zero return. Here it is with respect to the prediction from all other stocks. Any number of variants in how many stocks to trade and in what proportions are of course possible.

\begin{figure}[ht]
\centering
\includegraphics[width=16cm]{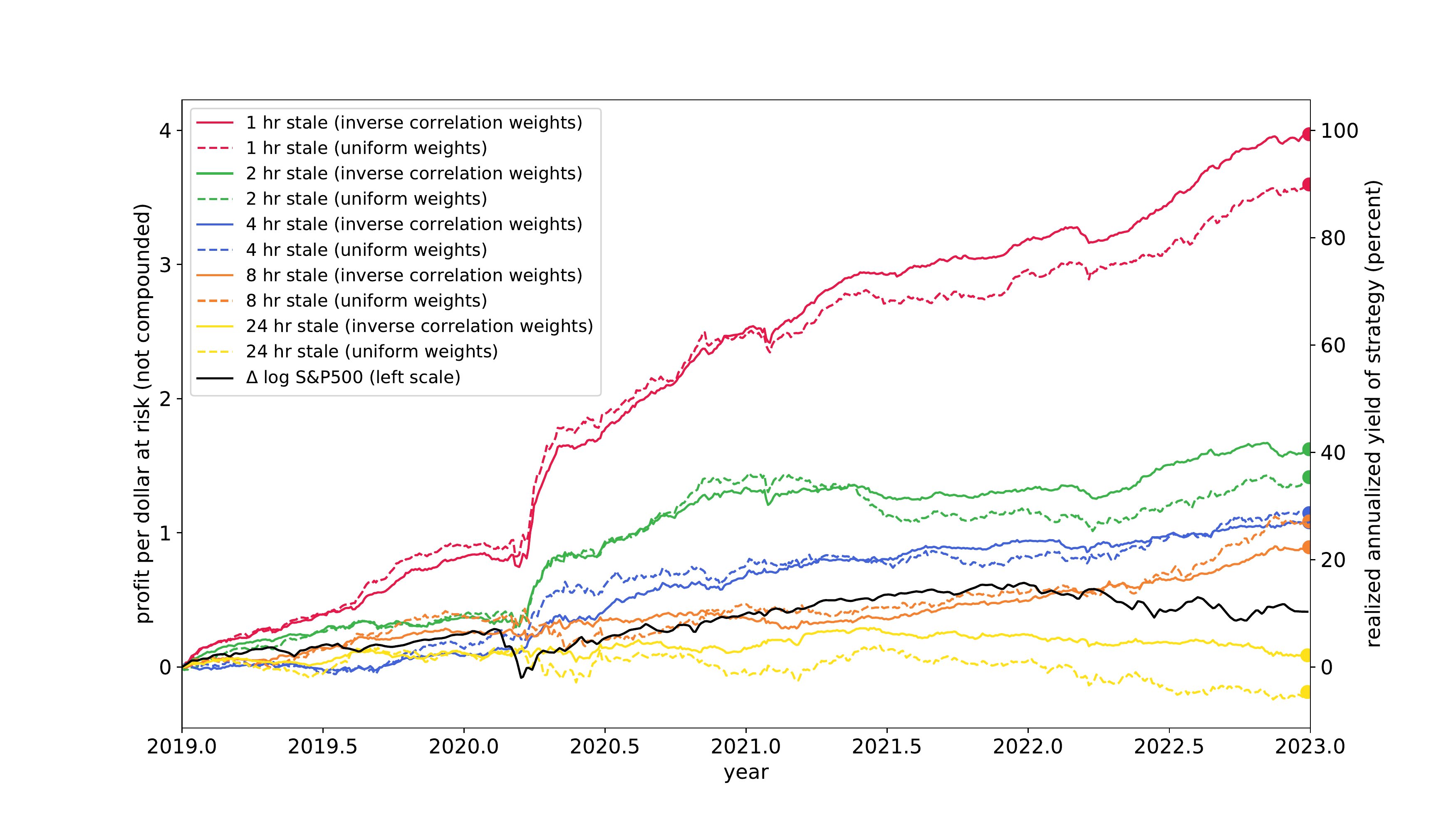}
\caption{Similar to Figure \ref{fig3}, growth of a unit stake (uncompounded) under an arbitrage trading strategy based on the discrepancy between a stock's hourly return and the return predicted by its correlation with all other stocks. Dashed lines use the naive prediction with all stocks equally correlated. Solid lines use measured correlations of earlier year 2018. Colors show different assuptions for how soon observed returns can be acted on (how stale is the data). Also shown is growth of the S\&P 500 index on the same scale. The right-hand scale shows the implied annualized yield of the end-point of each curve.}
\label{fig11}
\end{figure}

Figure \ref{fig11} shows the result for varying values of the staleness parameter, and for both the measured correlation and the naive assumption of all stocks equally correlated. Comparing to Figure \ref{fig3}, one sees better performance, measured as mean annualized yield, by about a factor two, yielding approximately 99\% annualized yield for the case of staleness $S=1$ (that is, trading in hour $h+2$, immediately after hourly returns are calculated from prices in hours $h$ and $h+1$).

The different colors in Figure \ref{fig11} show that the staleness matters greatly. Delaying trades by one or more additional hours causes a precipitous drop in profits. At 24 hours delay, the arbitrage strategy fails completely.  We see also that generally, but not always, using the measured correlations is superior to an unweighted average.

\section{Discussion}
\label{sec9}

That mean-reversion behavior on intraday timescales can be found in stock markets is not a new result. Indeed, ChatGPT-4, by design a font of conventional wisdom, opines that such microstructure noise can be due to spoofing, layering, overreactions, liquidity provision, profit-taking, order imbalances, and other kinds of bid-ask bounce \cite{chat}. What is perhaps surprising is that this mean-reversion is seen so continuously and systematically, over timescales from minutes to days, in $>1000$ frequently traded NYSE stocks (Figures \ref{fig2} and S2); that the effect is so close to a featureless Hurst exponential process with no characteristic timescale; and that it is, in backtesting with zero transaction costs, arbitrageable with annualized return on capital $\sim 50$\%.

That using the information of the full correlation matrix further improves these arbitrage rates of return is likewise not a surprise. A minor surprise is that most of the benefit accrues from a simplistic model in which all pairs of stocks have the same value $\rho$ (for any non-negligible numerical value), which leads to estimating any stock's ``expected" return as an unweighted average of the normalized returns of all other stocks. This simplistic model explains close to 40\% of stocks' hourly variances, as compared to about 50\% explained using the full correlation matrix. Also interesting is that the latter figure hardly depends on which single year in the range 2018-2022 is used to estimate the correlations, demonstrating a perhaps surprising persistence of correlational structure even during and across high-volatility COVID-19 year 2022.

We showed in \S\ref{sec62} that the observed, counterintuitive increase of stock correlations with time can be explained quantitatively by positing that the mean-reverting component of individual stock prices does not share the same correlation with other stocks as the random-walk component. That would imply that while, on average, stocks slightly overreact to news, they do so independently of one another, not coherently with the average market, a statistical model embodied in equation \eqref{eq19}. Alternative explanations might of course be possible.

\FloatBarrier

\printbibliography

\makeatletter
\renewcommand \thesection{S\@arabic\c@section}
\renewcommand\thetable{S\@arabic\c@table}
\renewcommand \thefigure{S\@arabic\c@figure}
\makeatother
\setcounter{section}{0}
\setcounter{figure}{0}

\section*{Supplementary Information}
\section{Further Detail of Calculations}
\subsection{Effect of Interval Averaging on Price-Difference Returns and Variogram Accuracy}
\label{suppinfo11}
We consider a memoryless Gaussian process. For simplicity, we calculate in the discrete-time case, that is, a random walk, then take the limit to a continuous process.

Suppose a time series of $2M$ normal deviates $r_i$, $i=1,\ldots,M$, $r_i \sim \Nor(0,1)$, representing microscopic returns. Their cumulative sum,
\begin{equation}
    p_k \equiv \sum_{i=1}^k r_i,\qquad k=1,\ldots,M
\end{equation}
represents a time series of prices.

We can calculate the macroscopic return over a time $M$ in two ways. The point-price method just differences the midpoints of the two consecutive intervals each of length $M$,
\begin{equation}
    R_{pp} = p_{3M/2} - p_{M/2}
\end{equation}
By the random-walk property, this has variance $\Var(R_{pp}) = M$.

Alternatively, we can calculate the average price $\overline{P}_0$ for the first $M$ points, $\overline{P}_1$ for the second $M$ points, and difference these.
\begin{equation}
    \overline{P}_0 = \frac{1}{M}\sum_{i=1}^M p_i
    = \frac{1}{M}\sum_{i=1}^M \sum_{j=1}^i r_j
\end{equation}
In the double sum, the term $r_j$ occurs $M-j$ times. This allows the immediate calculation of $\Var(\overline{P}_0)$ as
\begin{equation}
    \Var(\overline{P}_0) = \Var\left(\frac{1}{M}\sum_{k=1}^M k r_k \right)
    = \frac{1}{M^2} \sum_{k=1}^M \Var(k r_k) = \frac{1}{M^2}\sum_{k=1}^M k^2 \approx
    \tfrac{1}{3} M
\end{equation}
and similarly for $\Var(\overline{P}_1)$. Since $\overline{P}_0$ and $\overline{P}_1$ are independent Gaussian variables, the variance of their difference is
\begin{equation}
    \Var(\overline{P}_1 - \overline{P}_0) = 2 \Var(\overline{P}_0) \approx \tfrac{2}{3}M
\label{eqcheat}
\end{equation}
So the ratio of the two variances as $M \rightarrow \infty$ is 2/3, as stated in the main text. (Equation \eqref{eqcheat} seems a bit of a cheat, so we checked the overall result by simulation!)

\begin{figure}[ht]
\centering
\includegraphics[width=17cm]{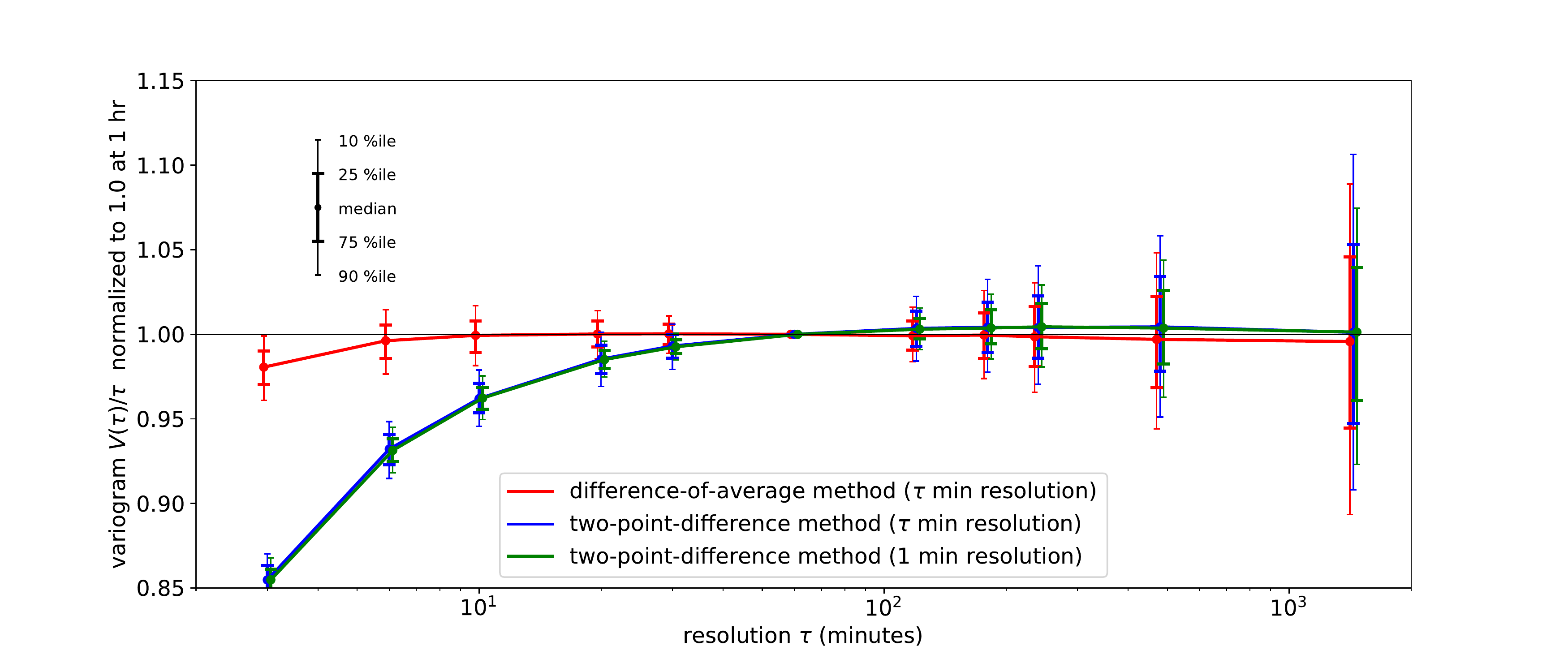}
\caption{Three methods for estimating the variogram $V(\uptau)$ applied to simulated one-minute price candles (open, high, low, close) in 1000 simulated one-year periods of memoryless Gaussian
returns. Each variogram is normalized to one at one hour, with quantile points shown for other
values of $\uptau$. For candle data, itself a kind of average, the difference-of-average method gives the most consistent results.}
\label{fig1a}
\end{figure}

\subsection{Effect of Interval Averaging on Price-Difference Returns and Variogram Accuracy}
\label{suppinfo12}

We next check how, with finite data, the measurement error in $V(\uptau)$ might differ for the two calculational methods above.
For this, we simulated 1000 one-year time spans of memoryless Gaussian returns at one-second intervals. In each one-minute interval, we calculated a candle of prices (open, high, low close). Now taking the candles as given data, we calculated $V(\uptau)/\uptau$, which should be constant, by three methods, using one year's data at a time: first, the difference-of-average method; second, the two-point difference method with point prices assumed known on a grid of spacing $\uptau$; third, the two-point difference method keeping one-minute resolution even for larger $\uptau$.

Figure \ref{fig1a} shows the results. Each one-year time span's variogram is normalized to one at $\uptau = 1$ hr, and the quantile points (median, etc.) of the 1000 years are shown for other values of $\uptau$. Dispersions increase with increasing $\uptau$ because the number of independent samples in a year decreases. The accuracies of all three methods are seen to be comparable, with method three a slight winner. However, both two-point difference methods show a bias that increases as $\uptau$ decreases towards one minute. This arises most likely because the candle prices are not point prices but themselves averages. 

\subsection{Relation Between Fraction of Variance Explained (FVE) and Fractional Mean Square Error (FMSE)}
\label{suppinfo13}
In the limited context of equations \eqref{eq29} and \eqref{eq31} above, we have measurements $r$ and estimates $\hat r$, both of zero mean and scaled to the same variance. Then,
\begin{equation}
    \text{FMSE} = \frac{\left< (\hat r - r)^2\right>}{\Var(r)}
    = \frac{2 \Var(r) - 2 \left<\hat r r \right>}{\Var(r)}
    = 2 - 2\rho
\label{eqSI13}
\end{equation}
where $\rho$ is the correlation $[\left<\hat r r \right>/\Var(r)]$. As is well known, the fraction of variance explained is $\rho^2$, so solving equation \eqref{eqSI13} for $\rho$ and squaring gives
\begin{equation}
    \text{FVE} = \rho^2 = \left( 1 - \tfrac{1}{2}\text{FMSE}\right)^2
\end{equation}

\FloatBarrier

\section{Observed Variance of NYSE Stocks in Five Years}
\label{secfiveyear}
Figure \ref{fig2} showed data from 2019 only. Here we show data for all years 2018--2022. Except for year 2020, results well fit by a small negative exponent $-0.07$. Year 2020 shows a similar trend, but exhibits some additional structure.
\begin{figure}[h!]
\centering
\includegraphics[width=14cm]{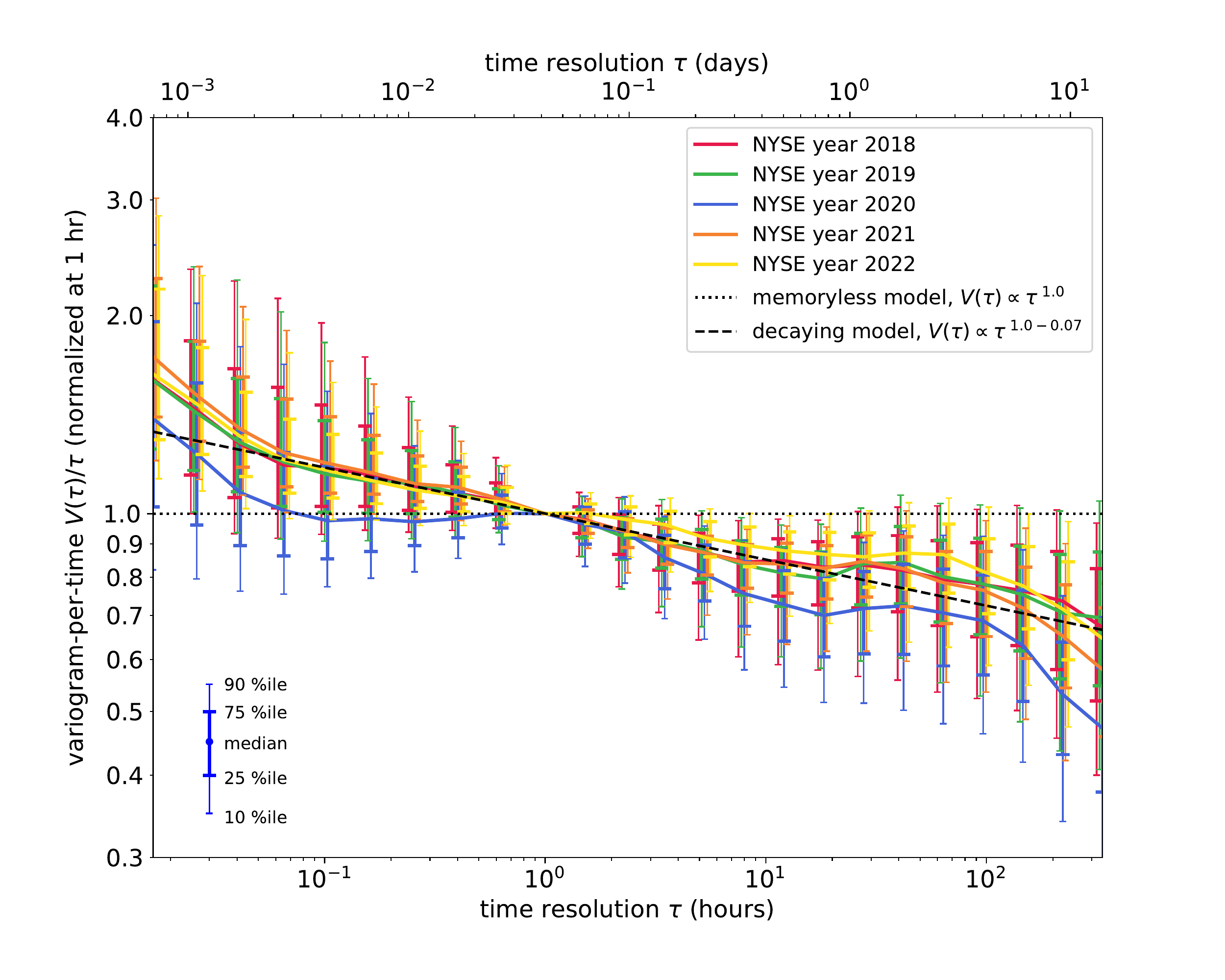}
\caption{Observed variogram $V(\uptau)$ for 1,091 NYSE stock prices as a function of resolution time interval $\uptau$ during the indicated 1-year periods. The ordinate is $V(\uptau)/\uptau$, so a memoryless random walk plots as constant. Except for year 2020, results well fit by a small negative exponent $-0.07$.}
\label{figS1}
\end{figure}

\FloatBarrier

\section{An Atlas of Correlation Embeddings for NYSE Stocks}
\label{secatlas}

\begin{figure}[h!]
\centering
\includegraphics[width=14cm]{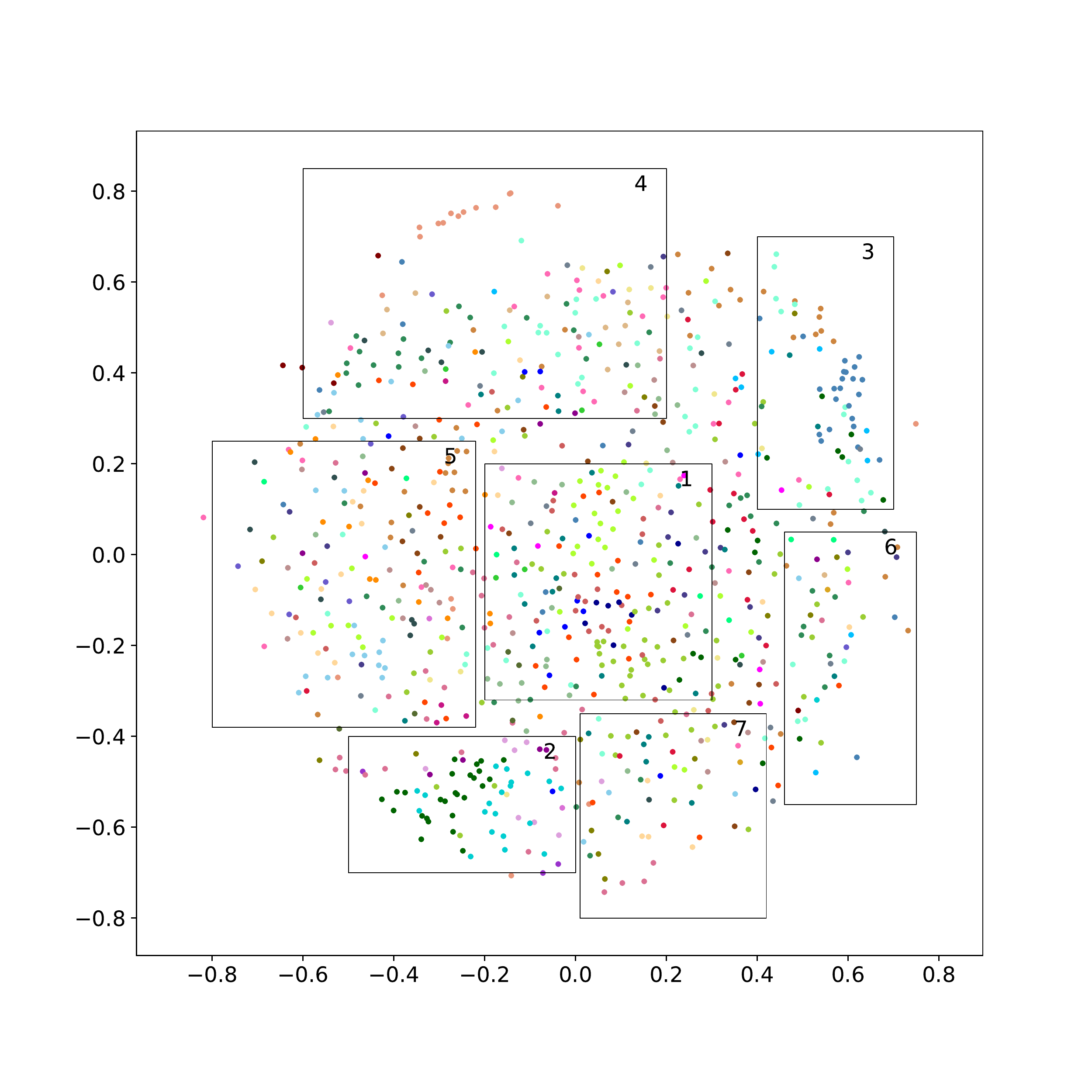}
\caption{The correlation coefficient $\rho_{ij}$ for hourly returns of all pairs $(i,j)$ of 866 NYSE stocks is computed. The distance between pairs is defined as $1-\rho_{ij}$. The resulting correlational structure is here projected into a two-dimensional space. Each dot is an individual stock, colored by its industry classification. The $x$ and $y$ axes each show the distance scale, but have no particular meaning (i.e., the figure could be arbitrarily rotated. The numbered boxes refer to subsequent Figures.}
\label{figAtlas0}
\end{figure}

\begin{figure}[ht]
\centering
\includegraphics[width=16cm]{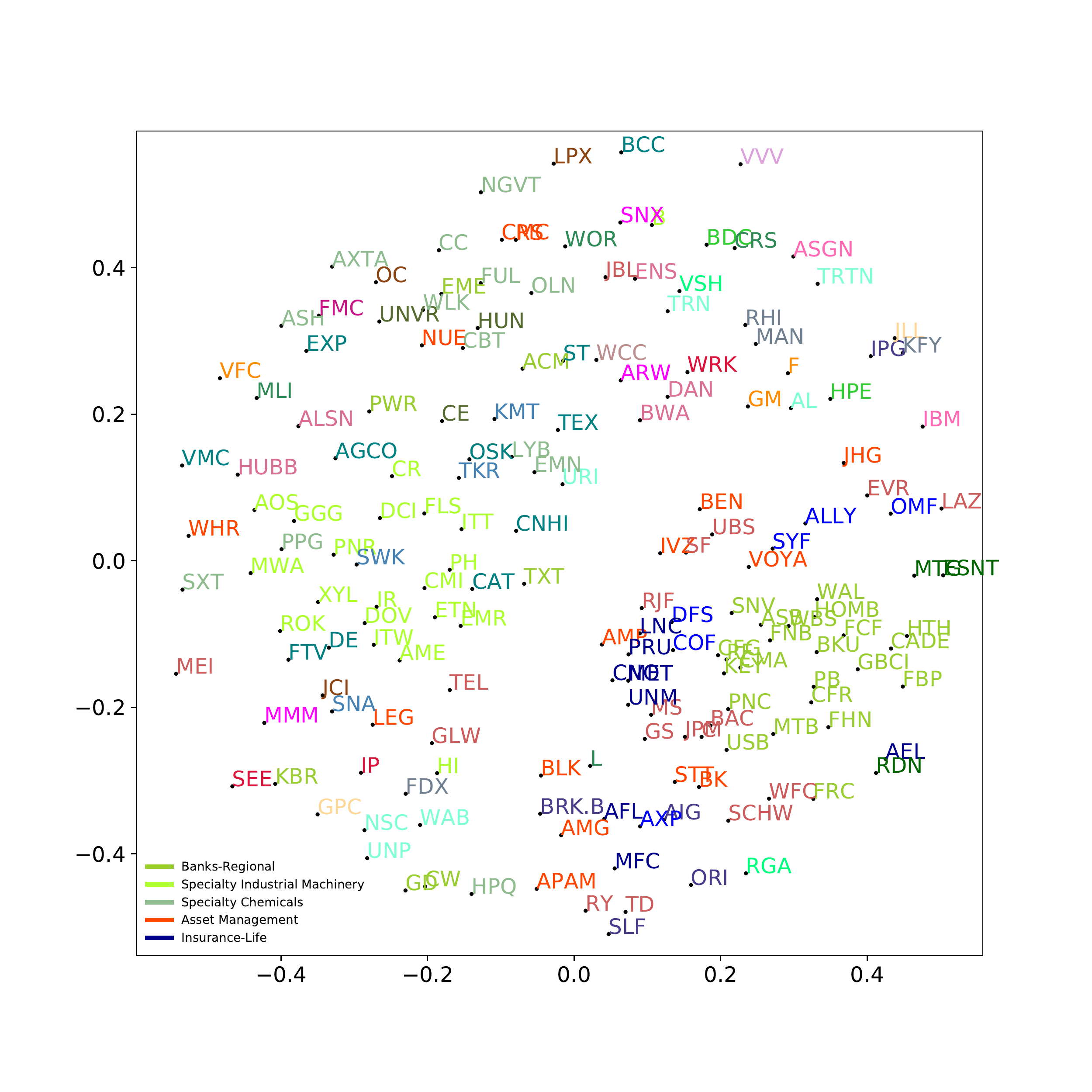}
\vspace{-1cm}
\caption{An embedding is re-computed for just the stocks in numbered box 1 in Figure
\ref{figAtlas0}, so as to better separate the industries in that box. Stocks are shown by their NYSE tickers and colored by their industry. (Inevitably there are chance collisions giving the same or similar colors to different industries.) The five most represented industries are named in the legend. In this projection, one sees a tight clustering of large (brown) and regional banks, closely flanked by life insurance, asset-management, and (orange) financial companies. On the other side of the figure, largely uncorrelated, one sees clusters of specialty chemicals and industrial machinery.}
\label{figAtlas1}
\end{figure}

\begin{figure}[ht]
\centering
\includegraphics[width=16cm]{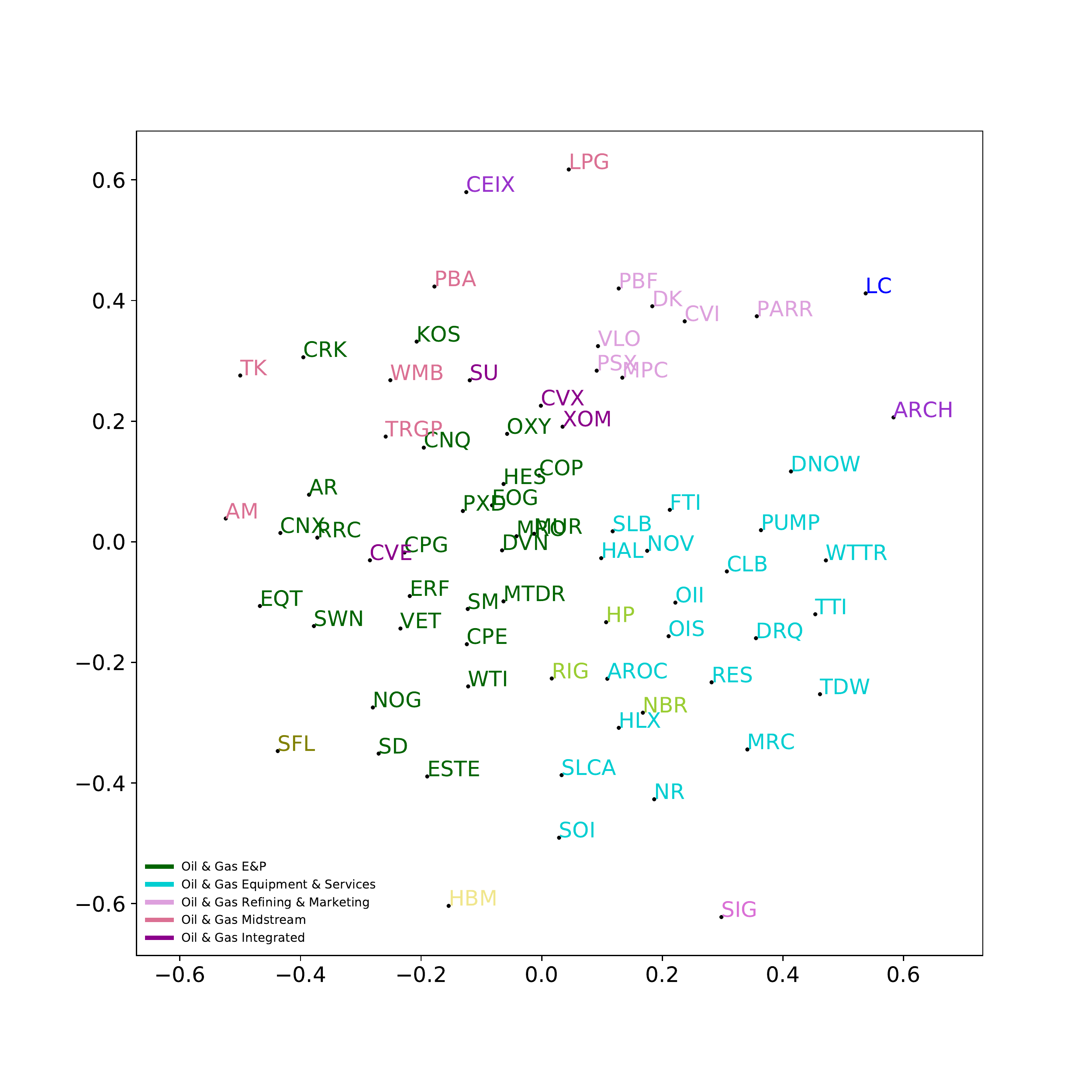}
\caption{Same as Figure
\ref{figAtlas1}, but for box 2 in Figure \ref{figAtlas0}. This region of correlation space is dominated by fossil fuel production, with subclusters of exploration and production, equipment and services, midstream, refining, etc.}
\label{figAtlas2}
\end{figure}

\begin{figure}[ht]
\centering
\includegraphics[width=16cm]{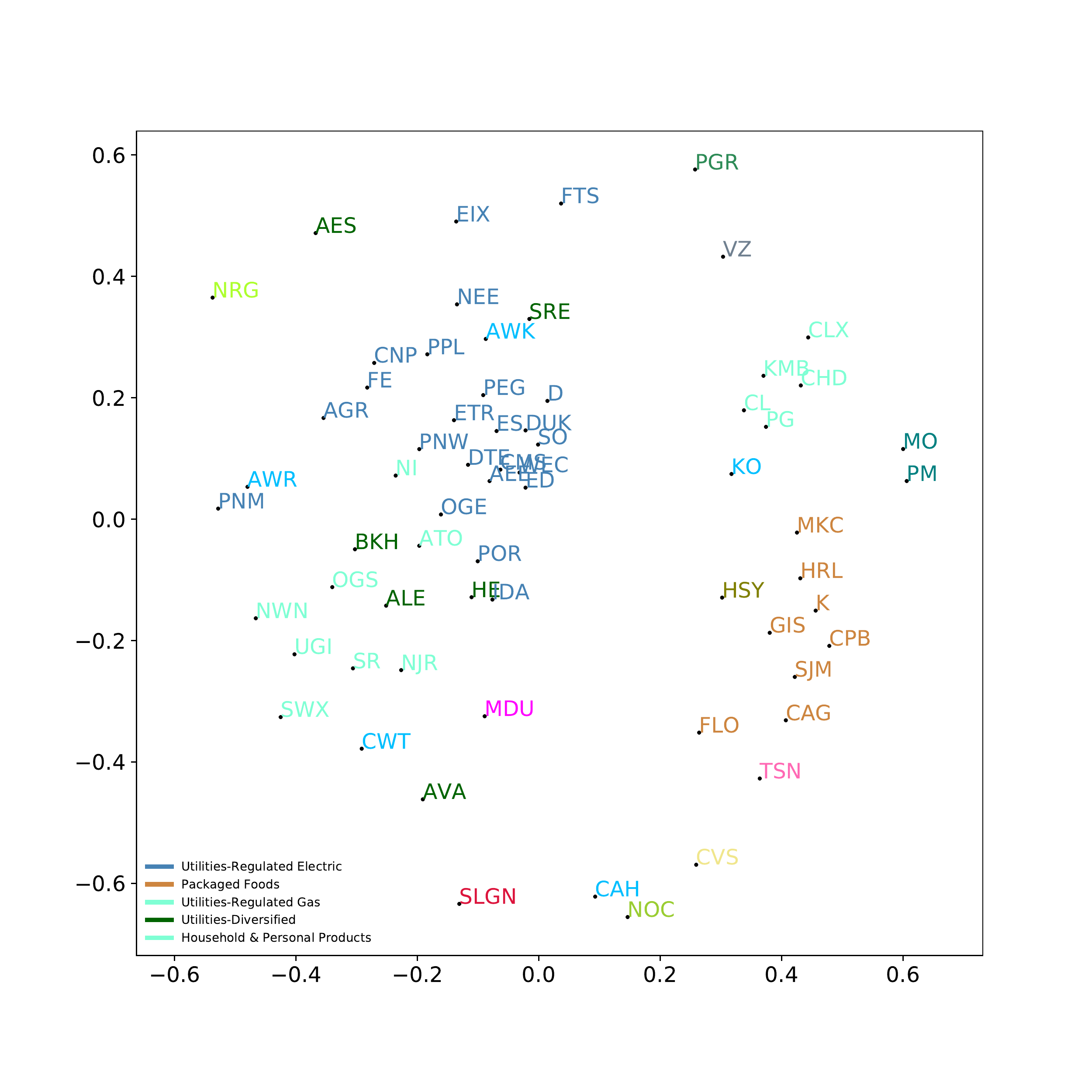}
\caption{Same as Figure
\ref{figAtlas1}, but for box 3 in Figure \ref{figAtlas0}. The figure's left has related clusters of utilities, while its right shows the related household and personal products and packaged foods.}
\label{figAtlas3}
\end{figure}

\begin{figure}[ht]
\centering
\includegraphics[width=16cm]{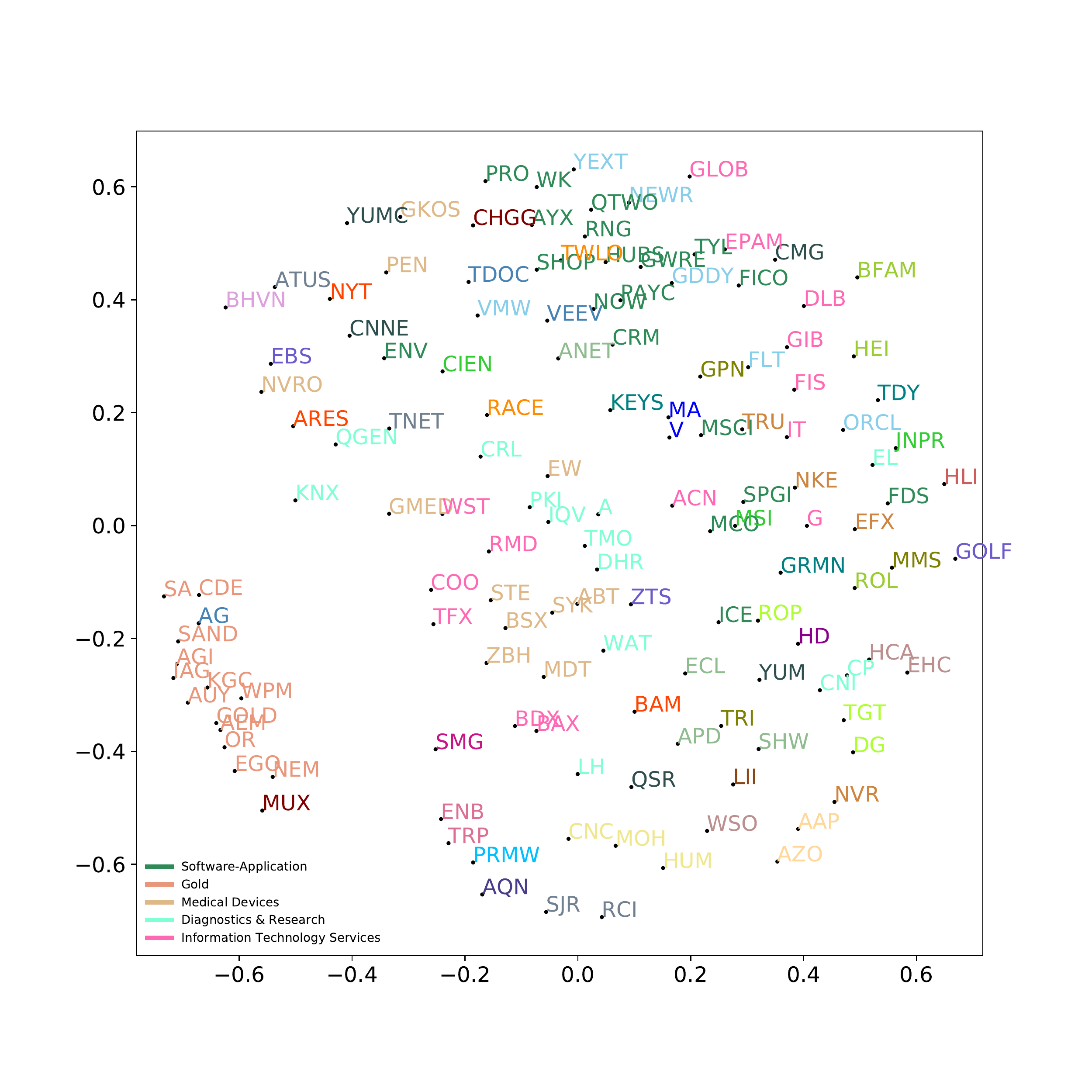}
\caption{Same as Figure
\ref{figAtlas1}, but for box 4 in Figure \ref{figAtlas0}. Gold stocks are prominent at the lower left. Software and IT cluster in the middle third, upper half.}
\label{figAtlas4}
\end{figure}

\begin{figure}[ht]
\centering
\includegraphics[width=16cm]{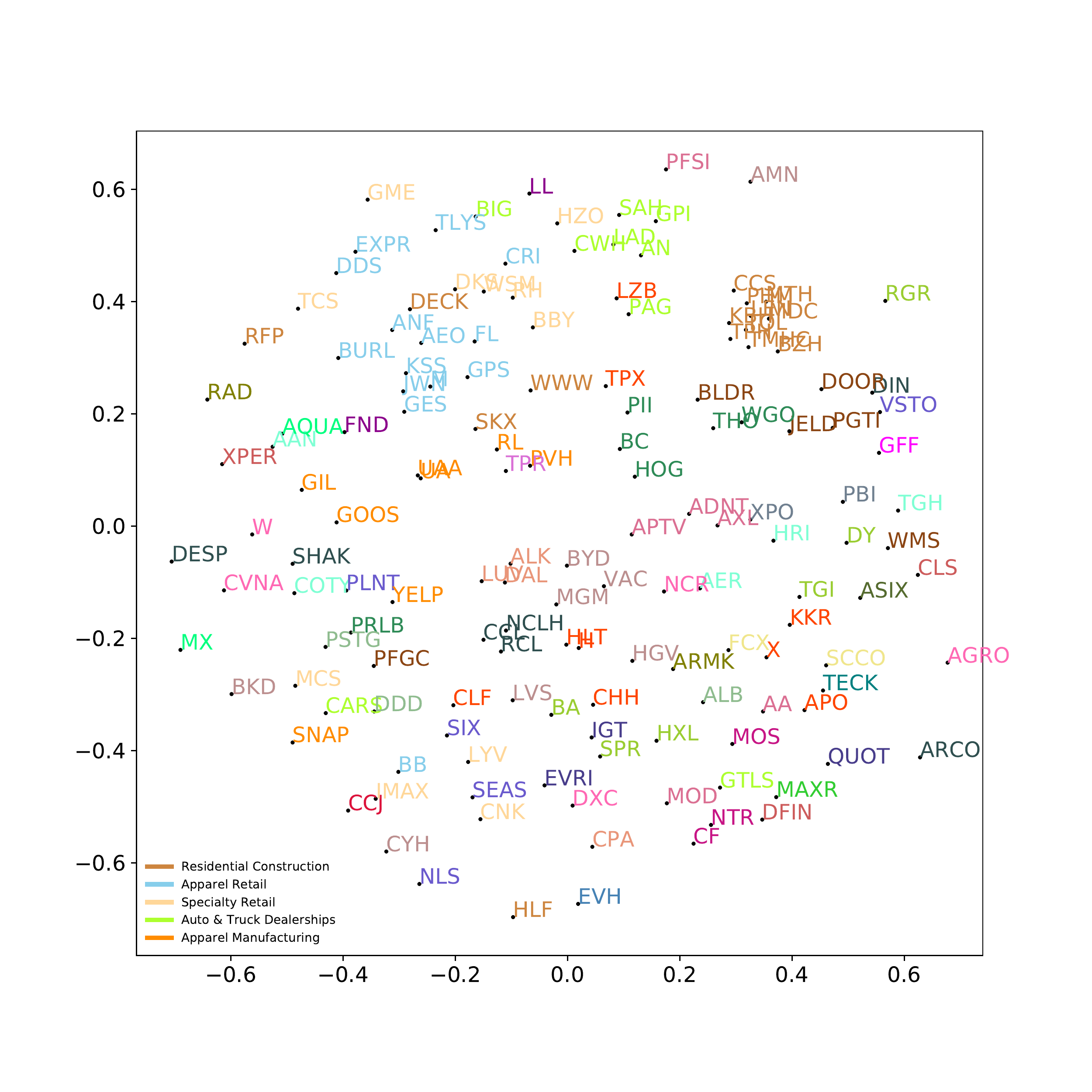}
\caption{Same as Figure
\ref{figAtlas1}, but for box 5 in Figure \ref{figAtlas0}. Residential construction and retail apparel form unrelated tight clusters. Cruise lines (dark gray) and gaming and resort (rose) are small clusters near the center, with most other industries quite spread out.}
\label{figAtlas5}
\end{figure}

\begin{figure}[ht]
\centering
\includegraphics[width=16cm]{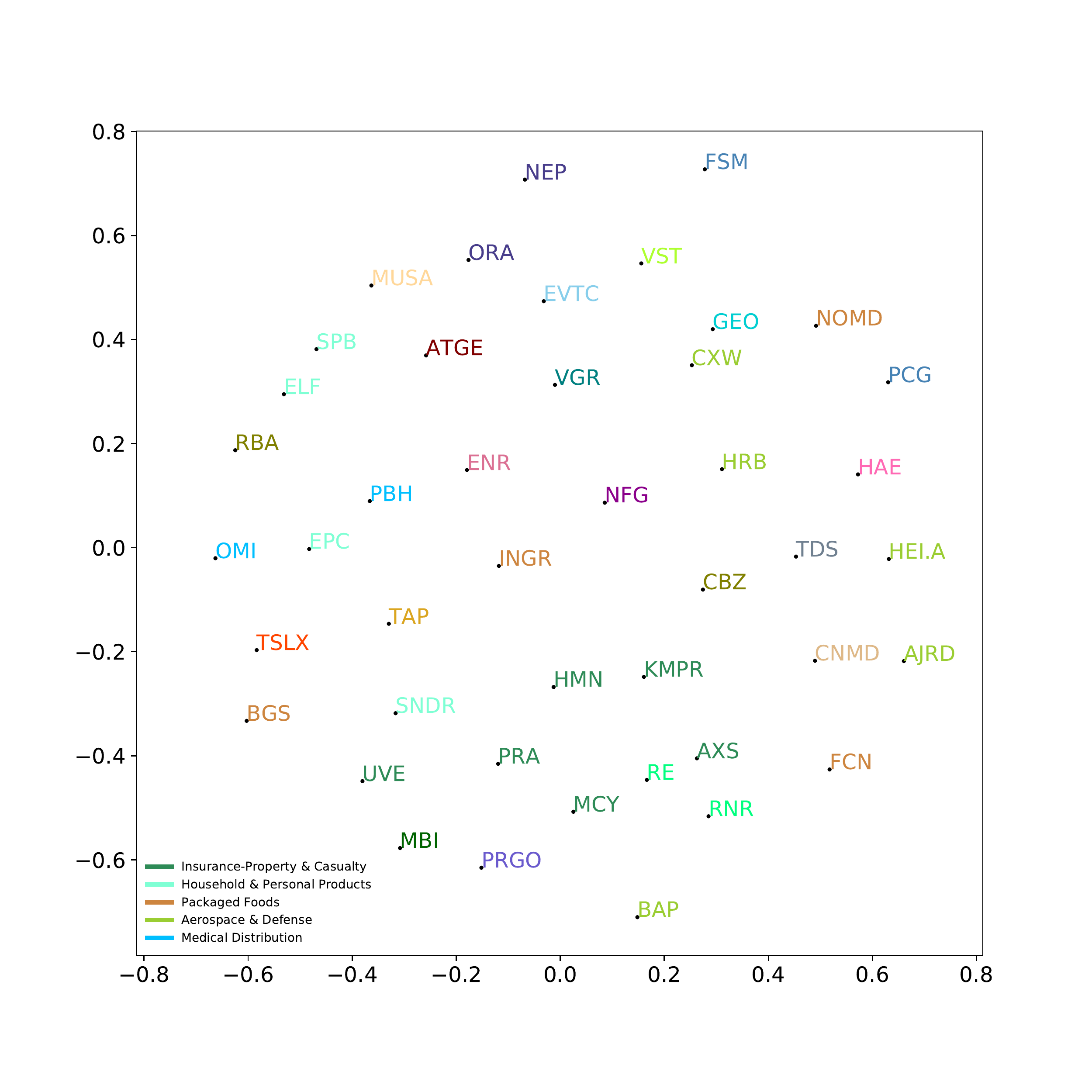}
\caption{Same as Figure
\ref{figAtlas1}, but for box 6 in Figure \ref{figAtlas0}. In this detail figure, and also the next, correlations are quite muted with individual tickers separated by distances implying $\rho \lesssim 0.8$ even for stocks in the same industry.}
\label{figAtlas6}
\end{figure}

\begin{figure}[ht]
\centering
\includegraphics[width=16cm]{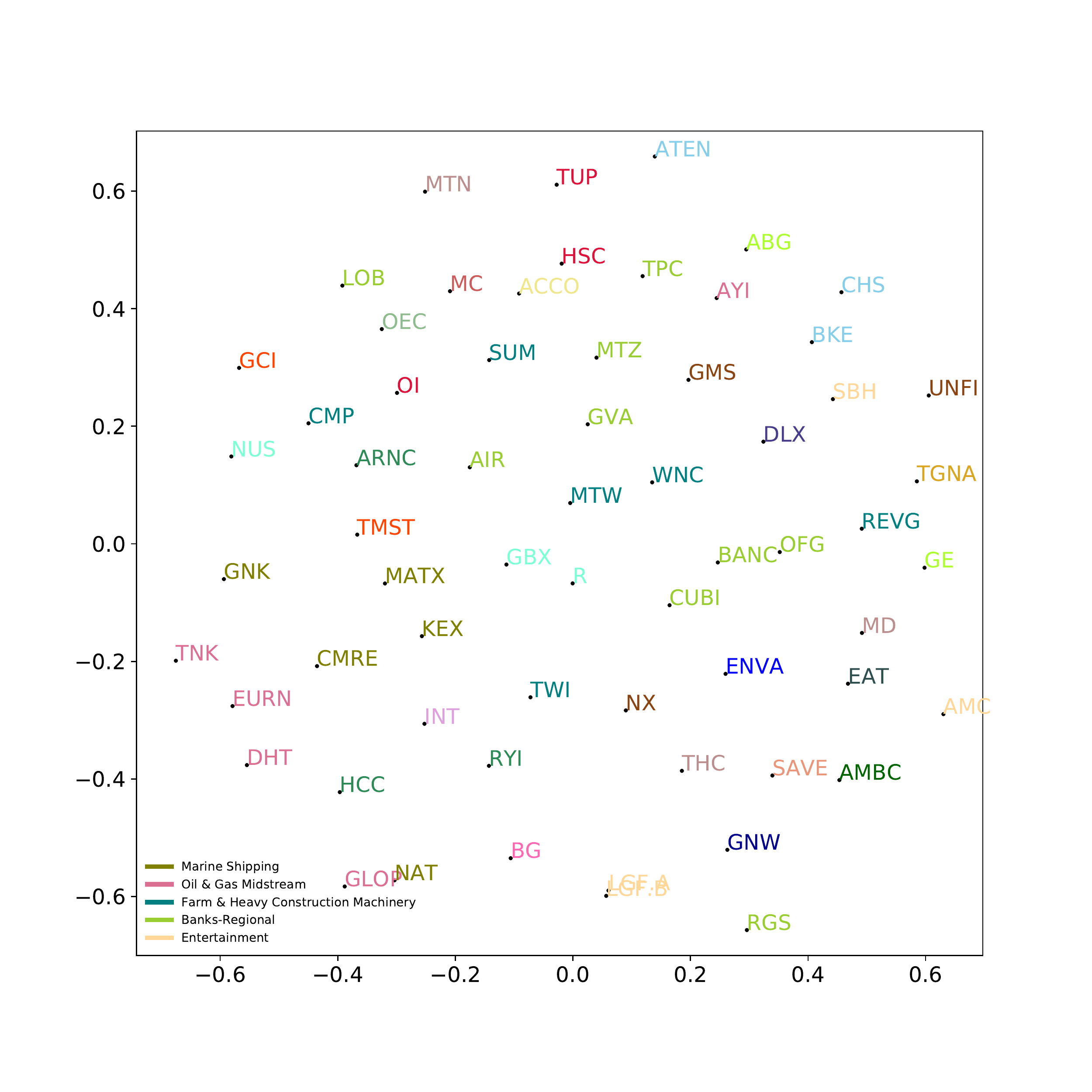}
\caption{Same as Figure
\ref{figAtlas1}, but for box 7 in Figure \ref{figAtlas0}.}
\label{figAtlas7}
\end{figure}

We constructed two-dimensional projections of the correlational structure of the 866 NYSE stocks that traded in all years, 2018--2022, using multidimensional scaling (MDS) \cite{MDSbook} as implemented in Scikit-Learn \cite{MDSsite}. (This differs slightly from the simplified description given in the main text.) The first figure following, Figure \ref{figAtlas0}, gives a top-level view of all the stocks at once. In this and subsequent figures the distance between any two points $i$ and $j$ is, insofar as the projection will allow, $1-\rho_{ij}$, where $\rho_{ij}$ is their pairwise correlation. It is important to understand that, because of the projection, clusters of points that are actually far apart may  sometimes be superposed; but, conversely, pairs that are far apart are genuinely little correlated. In Figure \ref{figAtlas0}, points are colored according to industry category, and the same colors are used in subsequent detail figures. In the detail figures (Figures \ref{figAtlas1}--\ref{figAtlas7}), the five most frequently occurring industries are identified in the legend. Individual stocks are identified by their NYSE ticker symbols, the lookup of which can easily identify their respective industries. Each detail figure is re-projected by a separate MDS so as to better separate less-correlated points, so it is not just a magnification of its box in Figure \ref{figAtlas0}.

\thispagestyle{empty}

\end{document}